\documentclass[fleqn,a4paper,authoryear,final,5p,times,twocolumn,10pt]{elsarticle}
\journal{J. Math. Phychol.}

\usepackage[breaklinks]{hyperref}
\usepackage[utf8]{inputenc}
\usepackage{graphicx}
\usepackage{color}
\usepackage{amsmath,amssymb,amsfonts,amsthm}
\usepackage{bm,bbm}

\newcommand{\be}{\begin{equation}}
\newcommand{\ee}{\end{equation}}
\newcommand{\ba}{\begin{eqnarray}}
\newcommand{\ea}{\end{eqnarray}}
\def\bs{\begin{subequations}}
\def\es{\end{subequations}}
\def\a{\alpha}
\def\b{\beta}
\def\de{\delta}
\def\g{\gamma}
\def\la{\lambda}

\def\De{\Delta}

\def\vr{\varrho}

\def\cA{\mathcal{A}}

\def\cC{\mathcal{C}}

\def\cF{\mathcal{F}}
\def\cG{\mathcal{G}}

\def\cS{\mathcal{S}}

\def\dc{d_\textsc{c}}
\def\dh{d_\textsc{h}}

\newcommand{\Eq}[1]{(\ref{#1})}
\def\com{\color{magenta}}
\def\cob{\color{blue}}
\renewcommand{\leq}{\leqslant}
\renewcommand{\geq}{\geqslant}
\newcommand{\boxd}[1]{\boxed{\phantom{\Biggl(}#1\phantom{\Biggl)}}}

\newcommand{\arX}[1]{\href{http://arxiv.org/abs/#1}{{\ttfamily\com arXiv:#1}}}
\newcommand{\doin}[6]{\href{http://dx.doi.org/#1}{{\cob {\it #2 #3} {\it #4}, #5}}}
\newcommand{\doinn}[5]{\href{http://dx.doi.org/#1}{{\cob {\it #2} {\it #3}, #4}}}
\newcommand{\doinnf}[5]{\href{http://dx.doi.org/#1}{{\cob\rm #2 #3 (#5) #4}}}

\newcommand{\ndoinn}[5]{\href{#1}{{\cob {\it #2} {\it #3}, #4}}}

\newcommand{\procsing}[7]{#2 (Ed.), \emph{#1} (pp.~#7). #4, #5: #3}
\newcommand{\procmany}[7]{#2 (Eds.), \emph{#1} (pp.~#7). #4, #5: #3}
\newcommand{\procmanys}[6]{#2 (Eds.), \emph{#1}. #4, #5: #3}
\newcommand{\procmanyd}[4]{#3 (Eds.), \href{#1}{\cob\it #2} (pp.~#4)}
\newcommand{\book}[5]{\emph{#1}. #3, #4: #2}
\newcommand{\tia}[1]{#1.}

\begin{document}

\begin{frontmatter}

\title{{\bf The geometry of learning}}
\author{Gianluca Calcagni}
\ead{g.calcagni@csic.es}
\address{Instituto de Estructura de la Materia, CSIC, Serrano 121, 28006 Madrid, Spain}

\date{May 2, 2016}

\begin{abstract}
We establish a correspondence between Pavlovian conditioning processes and fractals. The association strength at a training trial corresponds to a point in a disconnected set at a given iteration level. In this way, one can represent a training process as a hopping on a fractal set, instead of the traditional learning curve as a function of the trial. The main advantage of this novel perspective is to provide an elegant classification of associative theories in terms of the geometric features of fractal sets. In particular, the dimension of fractals can measure the efficiency of conditioning models. We illustrate the correspondence with the examples of the Hull, Rescorla--Wagner, and Mackintosh models and show that they are equivalent to a Cantor set. More generally, conditioning programs are described by the geometry of their associated fractal, which gives much more information than just its dimension. We show this in several examples of random fractals and also comment on a possible relation between our formalism and other ``fractal'' findings in the cognitive literature.
\end{abstract}

\begin{keyword}
Pavlovian conditioning \sep associative models \sep fractal geometry
\end{keyword}

\end{frontmatter}


\section*{Highlights}
\begin{itemize}
\item A correspondence between Pavlovian conditioning processes and fractals is proposed.
\item This duality is applied to many associative theories and conditioning programs.
\item $1/f$ scaling in human cognition and a random fractal model are compared.
\item Slow learning can be interpreted as an excitatory process contaminated by inhibition.
\item Individual response is characterized by progressively damped fluctuations.
\end{itemize}

\section{Introduction} 

Making psychology quantitative has been a difficult but feasible challenge since the first laboratory of psychophysiology established by Wundt. Making it a mathematical theory under analytic control has been, and possibly will always be, a utopia. Nevertheless, there are a plethora of analytic models which are able to fit and explain data coming from the observation of subjects in specific experiments. For instance, in the context of behavioral theories of Pavlovian conditioning, one can study the interplay between a conditioned stimulus (CS) and the subsequent occurrence of an unconditioned stimulus (US) of typically high relevance for the subject, such as food or an electric discharge. Despite their limited range of applicability, associative conditioning models are useful for several reasons. First, they express in a compact and economic way concepts that took time and many pages to be formulated. For example, the fact that ``the prior activity influences the value of'' the stimulus, recognized since the early stages of functionalism \citep{Dew96}, translates effectively in a description of conditioning as an iterative process progressively modifying the strength of the association and the salience of the stimulus. Second, they offer novel insights that can be easily checked and falsified quantitatively as new data and experimental designs become available. The question we would like to pose in this paper, limited to animal and human behavior, is: How much can we expand our toolbox of mathematical models in order to extract valuable information on learning processes?

The classic 1950s theoretical approaches to simple cases of conditioning are cast in the language of probability theory [see, e.g., the works by Bush and Mosteller \citeyearpar{BuMo1,BuMo2,BuMo3}, Estes \citeyearpar{Est50}, Estes and Burke \citeyearpar{EsBu1}, and the reviews by \citet{Bow94} and \citet{Mos58}]. In these models, one considers the probability $p$ of a given conditioned response (CR) as a function of the trial number $n$. The increment $\De p_n$ at each trial is linear in $p_n$; by evaluating $p_n$ iteratively, one obtains a learning curve. Alternatively, the probability $p$ can be replaced by the strength of association $V$. This change of variable is useful for phenomenological applications because $V$, although mediated by internal variables such as the organism's motivational state or attention, can directly be measured by several performance indicators, \emph{in primis} the subject response. For instance, the quality of surprise in the US as a function of the appearance of the CS was first suggested by Kamin in relation with cue competition (Kamin, \citeyear{Kam68,Kam69}). For a single CS, the evolution of the novelty (or ``surprisingness'') of the US along the learning curve had been made quantitative already by Hull in his linear model of Pavlovian conditioning \citep{Hul43}. Recast in modern terminology by Rescorla and Wagner \citeyearpar{RW72} and Wagner and Rescorla \citeyearpar{WR72}, this model states that the change $\De V_n$ in the strength of the association at the $n$th trial is
\be\label{DeV}
\De V_n=\a\b(\la-V_{n-1})\,,\qquad n=1,2,3,\dots,
\ee
where $0\leq\a\leq 1$ is the salience of the CS, $0\leq\b\leq 1$ is the salience of the US, and $0\leq\la\leq 1$ is the magnitude or intensity of the US (i.e., the asymptote of learning). The term $\la-V_{n-1}$ indicates the surprisingness of the US, which decreases as the associative strength increases. The association strength gained up to the start of the $n$th trial can be found iteratively:
\be\label{rewam}
V_n=V_{n-1}+\De V_n=(1-\a\b)V_{n-1}+\a\b\la\,.
\ee
The solution of this equation is
\be\label{lacu1}
V_n=\la[1-(1-\a\b)^{n-1}]\,.
\ee
When no association has been made yet, at the beginning of the first trial $V_1=0$, which fixes the unphysical constant $V_0=-\a\b\la/(1-\a\b)$. When $\la\neq 0$, the conditioning is excitatory and the US always occurs after the CS. Maximum learning is achieved when $V=\la$. If $\la=0$, the US does not show up after the CS and the conditioning is inhibitory or of extinction. Rescorla and Wagner extended the linear model to the case of the presentation of multiple CSs \citep{RW72,WR72}, as we will discuss later.

The main contribution of this paper is to propose a geometric interpretation of learning processes which carries several advantages. First, it is useful at the time of assessing the efficiency of these processes quantitatively, both within a given model (how efficiency is affected by the salience of the stimuli for the subject) and when comparing different models. The \emph{efficiency} of an excitatory conditioning can be roughly defined as the inverse of the number of trials necessary to increase the associative strength from 0 to, say, $0.9\,\la$. This concept is subject-dependent and may be used either to compare the learning of different individuals within the same program or, when averaging over individuals within the same experimental group, to compare different learning programs.

Specifically, we obtain the following results. (i) We recognize Eq.\ \Eq{rewam} as one of the similarity maps defining Cantor sets, which are an example of peculiar, totally disconnected sets known as deterministic fractals. (ii) We calculate the Hausdorff dimension $\dh$ of the set for Hull's model and show that it depends on the parameters $\a$ and $\b$ in such a way that \emph{the smaller the dimension, the more efficient the conditioning}. (iii) This picture can be generalized to any other conditioning described by iterative equations, giving explicit multidimensional examples that include Rescorla--Wagner, Mackintosh, and Pearce--Hall models. As a further application to nonlinear sets, (iv) we approximate Mackintosh theory (in the case of a single cue) with a new model where the recursive equation describes slow learning at intermediate trials; the dimension of this conditioning process is calculated and shown to be greater than in the Hull model for the same asymptotic value of the parameters, in agreement with (ii). Note that, in the presence of a single cue, the learning rate is already enough to compare different individuals or programs. One can see this by noting that the Hausdorff dimension \Eq{dc2} only depends on the product of the saliences and the smaller the salience, the larger the dimension. Nevertheless, when one goes beyond single-cue configurations and considers more complicated settings (Section \ref{general}), it may become progressively difficult or ambiguous to define effective learning rates. On the contrary, the Hausdorff dimension is always a well-defined parameter that provides a quick way to compare different individuals or models. Unfortunately, in practice, calculating the Hausdorff dimension for complicated deterministic processes may be as difficult as deciding on effective learning rates. However, the fractal paradigm is not limited to the definition of a new parameter, and its advantages do not end here. (v) The rethinking of learning processes in geometric terms will allow us to reinterpret conditioning as a \emph{mixture of excitatory and inhibitory processes} rather than a black-or-white selection of either. The degree of mixing will be determined by the value of the Hausdorff dimension (Section \ref{conseq}). 

(vi) Also, we generalize the construction to random fractals, which are essential to describe experimental designs of Pavlovian conditioning where the characteristic of the stimuli are determined by random algorithms or the US is not presented at all trials (partial reinforcement). The Hausdorff dimension of the Cantor set is independent of the US intensity and it does not fully capture the efficiency of a process. This is obvious from Eq.\ \Eq{dc2} but (vii) we also give the counterexample of a partial-reinforcement program (know to be ``less conditioning'' than continuous reinforcement), where $\la=0$ in some of the trial but $\a$ and $\b$ (hence, $\dh$) are kept fixed throughout. Here the efficiency (the Hausdorff dimension, a pointwise geometric indicator) is less important than the determination of the geometric shape of the fractal, which offers a more global and useful perspective than the number $\dh$. In fact, (viii) the mappings generating the fractal give a prediction on the learning curve: there will be plateaux in the curve with such and such distribution determined by the random algorithm employed to pick the value of the parameters at each trial. Different randomizations of Hull's model will illustrate the point. Finally, (ix) we make a preliminary connection with some results in the cognitive literature on performance variability, which was found to follow a multifractal pattern. With all due caution in comparing widely different paradigms, we simulate performance variability of internal origin by a Pavlovian conditioning model where the salience of the stimuli slightly changes at each trial, according to a random algorithm. Since $\dh=\dh(\a\b)$ only depends on ``internal'' parameters determined by the type of subject and the type of stimuli presented, under a cognitive-interactionist perspective the Hausdorff dimension can be reinterpreted as the part of the efficiency of the process due to the characteristics of the subject in relation to the stimuli presented. Again, fractal geometry has the potential to open a new door of analysis.

The plan of the article is as follows. In Section \ref{fracan}, we recall some basic aspects of deterministic fractals. In Section \ref{geoin}, we apply this formalism to Hull's associative model of Pavlovian conditioning. Section \ref{general} is devoted to the generalization of this one-dimensional case to more realistic models with many cues or deterministically varying parameters (CS salience, US magnitude), such as Rescorla--Wagner (Section \ref{rwmod1}), Mackintosh and Pearce--Hall (Section \ref{rwmod2}), and a new nonlinear model akin to Mackintosh (Section \ref{nolin}). Random fractals are the subject of Section \ref{rafra}; flexible conditioning programs are discussed in Section \ref{rafra1}, where the fractal construction is extended to the very important case of random sets; a digression on cognitive experiments unveiling a multifractal pattern in task performance variability and its possible relation with our findings is discussed in Section \ref{relp}. Section \ref{conseq} briefly explores some applications of the fractal picture, both to the practical understanding of conditioning processes and to experimental predictions about response variability. Conclusions and future directions are in Section \ref{concl}.


\subsection{Main message for psychologists}

The presentation is rather mathematical since it borrows a few concepts from fractal geometry \citep{Fal03}. This may considerably disorient part of the target readership of this paper, namely, psychologists with a limited mathematical background. For this reason, we summarize here and in human language the core message of this work.
\begin{itemize}
\item Models of learning can be looked upon in a unified way. Under a change in perspective with respect to traditional presentations, all influential models of Pavlovian conditioning such as Hull's, Rescorla--Wagner's, and Mackintosh's can be understood as deterministic fractal sets. In particular, the dimension \Eq{dc2} of the fractal set relates to the efficiency (learning speed) of the associated conditioning model. This perspective links the somewhat scientifically isolated elementary learning models to the robust framework of fractal geometry and illustrates the potential for cross-disciplinary fertilization in psychological science.
\item Within the same framework, all deterministic conditioning models (which uniquely predict a certain association strength and stimulus saliences at any given trial) can be generalized to random models where the learning rate (saliences) and, consequently, the subject response fluctuate from trial to trial and one can only make probabilistic predictions. We will construct several examples. The key point is that, when regarded as a fractal, learning processes with a random component do not generate all possible values of associative strength, but only a subset of values. This subset is a random fractal. Efficient and fast conditioning is characterized as a ``discontinuous'' process with long hops between one point in the acquisition curve and the next. Under these conditions, inhibition is a completely separated process which does not interfere with acquisition. On the other hand, there seems to be a bit of inhibition in slow excitatory conditioning. This statement will be made mathematically precise.
\item From the point of view of empirical research, the reformulation of known conditioning models as deterministic fractals does not add new predictions. In this respect, reordering thoughts in terms of fractal geometry may interest the theoretician and the epistemologist, but leave the experimentalist skeptical about its practical usefulness. However, the fractal interpretation of random models does make a characteristic prediction about response variability (Section \ref{conseq}), which must be limited to the values of associative strength included in the fractal set. Concretely, we expect a specific general trend of response variability for a subject undergoing a training program equivalent to a random fractal, where the US is not present at all trials or the CS has variable salience. The learning curve in these two situations is depicted, respectively, in Figs.\ \ref{fig4} and \ref{fig7}. These features can be easily checked, even qualitatively, in laboratory experiments and the present work lays the theoretical ground for such a test.
\end{itemize}


\section{Fractals and the Cantor set}\label{fracan}

Let
\be\label{contr}
\cS_1(x)=a_1 x+b_1\,,\qquad \cS_2(x)=a_2 x+b_2\,,
\ee
be two \emph{similarity maps}, where $0<a_{1,2}<1$ (called similarity ratios) and $b_{1,2}$ (called shift parameters) are real constants and $x\in I$ is a point in the unit interval $I=[0,1]$. The rationale behind the term ``similarity'' (not to be confused with other usages in psychology) will be explained shortly. The image $\cS_i(A)$ of a subset $A\subset I$ is the set of all points $\cS_1(x)$ where $x\in A$. A \emph{Cantor set} or Cantor dust $\cC$ is given by the union of the image of itself under the two similarity maps \Eq{contr}, $\cC=\cS_1(\cC)\cup \cS_2(\cC)$. For instance, the ternary (or middle-third) Cantor set $\cC_3$ \citep{Can83} has $a_1=1/3=a_2$, $b_1=2/3$, and $b_2=0$:
\be\label{comap}
\cS_1(x)=\tfrac13 x+\tfrac23\,,\qquad \cS_2(x)=\tfrac13 x\,.
\ee
The above definition of $\cC$ is implicit but there exists also an explicit definition: letting $S(I):=\cS_1(I)\cup\cS_2(I)$ be the transformation on the interval $I$ (the standard symbol ``$:=$'' indicates that the left-hand side is defined by the right-hand side) and being $S^n=S\circ S\circ\cdots\circ S$ the $n$th iterate of $S$ (i.e., $S$ applied $n$ times), then $\cC=\cap_{n=0}^{+\infty} S^n(E)$. It can be shown that these two definitions are equivalent and generalizable to an arbitrarily large but finite number of maps $\cS_k$ \citep{Fal03} (see also Section \ref{general}). Moreover, for this generic iterated function system the resulting set $\cF$ (a deterministic fractal) always exists and is a unique attractor.

\begin{figure}
\centering
\includegraphics[width=7.8cm]{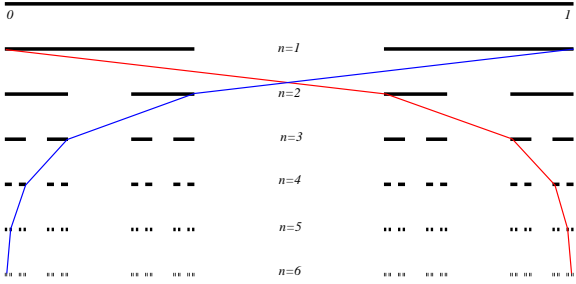}
\caption{\label{fig1} The ternary Cantor set for $n=6$ iterations of the maps \Eq{comap}. Interpreted as a representation of the Hull model of Pavlovian conditioning, the line $[0,1]$ is the range of possible values of the associative strength $V$ developed by the subject and $n$ is the trial number. The thin red curve is the learning curve connecting the points $V_n=0,2/3,8/9,26/27,\dots$ and it corresponds to an excitatory conditioning process where $V_1=0$ (no association at the start of the first trial) and $V_n$ tends to $\la=1$ progressively. The thin blue curve corresponds to inhibition or extinction, where $V_1=1$ and $V_n$ tends to $\la=0$. In the text, we discuss also conditioned inhibition. (For interpretation of the references to color in this figure legend, the
reader is referred to the web version of this article.)}
\end{figure}

The Cantor set $\cC$ is shown in Fig.\ \ref{fig1}. At the first iteration, the interval $[0,1]$ is rescaled by $1/3$ and duplicated in two copies: one copy (corresponding to the image of $\cS_2$) at the leftmost side of the unit interval and the other one (corresponding to $\cS_1$) at the rightmost side. In other words, one removes the middle third of the interval $I$. In the second iteration, each small copy of $I$ is again contracted by $1/3$ and duplicated, i.e., one removes the middle third of each copy thus producing four copies 9 times smaller than the original; and so on. Iterating infinitely many times, one obtains $\cC_3$, a dust of points sprinkling the line. The set is \emph{self-similar} inasmuch as, if we zoom in by a multiple of 3, we will observe exactly the same structure. Thus is explained the name ``similarity'': the maps $\cS_k$ make smaller copies of the set which are identical to the original except for their relative size.

It is easy to determine the dimensionality of the Cantor set $\cC$. Since this dust does not cover the whole line, it has less than one dimension. Naively, one might expect that the dimension of $\cC$ is zero, since it is the collection of disconnected points (which are zero-dimensional). However, there are ``too many'' points of $\cC$ on $I$ and, as it turns out, the dimension of the set is a real number between 0 and 1. In particular, given $N$ similarity maps all with ratio $0<a<1$, the similarity dimension or \emph{capacity} of the set is
\be\label{dc}
\dc(\cC):=-\frac{\ln N}{\ln a}\,.
\ee
This formula is valid for an exactly self-similar set made of $N$ copies of itself, each of size $a$. Note that $a= N^{-1/\dc}$: the smaller the size $a$, the smaller the copies at each iteration and the smaller the dimensionality of the set. In the case of the middle-third Cantor set, $N=2$ and $a=1/3$, so that $\dc=\ln2/\ln 3\approx 0.63$. Sets with noninteger dimensionality are called \emph{fractals}, a term coined by \citet{Man67}. There are other important geometric indicators used in fractal geometry, such as the box-counting dimension\footnote{For a set $\cF$ embedded in a $D$-dimensional space, the box-counting dimension is $d_\textsc{b}:=-\lim_{\delta\to 0} \ln N(\delta)/\ln \delta$, where $N(\delta)$ is the minimum number of $n$-balls, $n\leq D$ with radius $\delta$ (or $n$-cubes of edge length $\delta$; the choice of the covering set is irrelevant) centered at points in $\cF$ and such that they cover $\cF$ (i.e., each point in $\cF$ lies in at least one ball). The number $N(\delta)$ increases as $\delta$ decreases, approaching the behavior $N(\delta)\sim \delta^{-d_\textsc{b}}$ as $\delta\to 0$. Intuitively, a set with many irregularities requires more balls for being covered, and their number increases faster than expected; an example from the real world is an irregular porous surface, for which $d_\textsc{b}>2$ \citep{PA}. On the other hand, a surface with ``too many holes'' may require less balls than a smooth one.} and the \emph{Hausdorff dimension} $\dh$. For the class of fractals we will consider here, they are equal to the capacity \citep{Fal03}. The Hausdorff dimension\footnote{To measure the geometry of a set $\cF\subset\mathbb{R}^D$, one can extend the same idea of the box-counting dimension to the case of a minimal covering of $\cF$ with covering sets of different size. Let $|U|={\rm sup}\{\Delta(x,y):x,y\in U\}$ be the diameter of a set $U\subset\mathbb{R}^D$, i.e., the greatest distance $\Delta(x,y)$ between two points in $U$. A $\delta$-cover of $\cF$ is a countable or finite collection of sets $\{U_i\}$ of diameter at most $\delta$ that cover $\cF$: $\cF\subset \bigcup_i U_i$, with $0\leq |U_i|\leq\delta$ for all $i$. If $s\geq0$ is a real non-negative parameter, one can define $\vr^s_{\rm H}(\cF) := \lim_{\delta\to 0}{\rm inf}\left\{\sum_i|U_i|^s~:~\{U_i\}~\textrm{is a $\delta$-cover of $\cF$}\right\}$. This limit exists (it can also be 0 or $+\infty$) and is a measure, the $s$-dimensional Hausdorff measure of $\cF$. The Hausdorff measure obeys the scaling property $\vr^s_{\rm H}(\la x)=\la^{s}\vr^s_{\rm H}(x)$, where $\la>0$ is the scale factor of a dilation $x\to \la x$. One can show that $\vr_{\rm H}^s$ is nonincreasing with $s$ and there exists a critical value of $s$ at which the measure jumps from $+\infty$ to 0. This is the Hausdorff (or Hausdorff--Besicovitch) dimension of $\cF$ \citep{Hau18}: $\dh(\cF) := {\rm inf}\{s~:~\vr^s(\cF)=0\}={\rm sup}\{s~:~\vr^s(\cF)=+\infty\}$. The Hausdorff measure diverges for $s<\dh$, is zero for $s>\dh$, and can be 0, $+\infty$, or finite at $s=\dh$. Roughly speaking, the Hausdorff dimension is the scaling of the volume of the covering sets with respect to their linear size. More details can be found in \citet{Fal03}.} is one of the most popular among the fractal dimensions and can be calculated for sets [or even spacetimes! See, e.g., \citet{trtls} and references therein] much more general than those we will consider here. For this reason, we will often refer to $\dh$ together with (or rather than) $\dc$. Thanks to the equivalence $\dh=\dc$ for the Cantor set \citep{Fal03}, we will only need the definition \Eq{dc} in what follows, thus avoiding the delicate technicalities involved in $\dh$.


\section{Geometric interpretation of the Hull and Rescorla--Wagner models} \label{geoin}

Let us now apply these results, which are quite standard in fractal geometry, to simple cases of the Hull and Rescorla--Wagner models: (a) excitation, (b) extinction, and (c) inhibition. We consider two conditioning experiments, one where the CS is an excitatory stimulus ($\la\neq 0$, the US always follows the CS) and another where the same stimulus is inhibitory ($\la=0$, the US never follows the CS). For instance, the CS can be a sound or a light and the US food or a discharge. Interpreting the association strength $V_n$ as the point in the $n$th iteration of a set and comparing Eqs.\ \Eq{rewam} and \Eq{contr}, we see that the similarity ratio and shift parameter in the Hull model (one CS) is
\be\label{conpa}
a_1=a_2=a=1-\a\b\,,\qquad b_1=\a\b\la\,,\qquad b_2=0\,.
\ee
The images $\cS_1(I)$ and $\cS_2(I)$ correspond to the set of association strengths measured in, respectively, excitatory and inhibitory conditioning. We can pair the Hull and Rescorla--Wagner models with a Cantor set $\cC=\cS_1(\cC)\cup \cS_2(\cC)$ with parameters \Eq{conpa}. In the example of Fig.\ \ref{fig1}, $1-\a\b=1/3$, which can be obtained with $\b=1$ (maximum salience of the US) and $\a=2/3\approx 0.66$ (a highly salient CS). 

The actual excitatory conditioning process (a) is shown by the red learning curve in Fig.\ \ref{fig1} [touching upon the $\cS_1(I)$ branch with initial condition $V_1=0$], while the extinction curve is shown in blue [touching the $\cS_2(I)$ branch with initial condition $V_1=1$]. In the case of extinction (b), we can consider a two-phase experiment. The first phase is excitatory with a given CS, corresponding to the red curve. Then, the $n=1$ point of the blue curve corresponds to the first trial in a second phase of training where the association strength $V^{\rm CS}$ relates the absence of the US ($\la=0$) with the same CS used in phase 1, so that $V^{\rm CS}$ decreases from 1 to zero. In the case of an inhibitory conditioning process (c), the red curve of phase 1 describes the association strength $V^{\rm CS1}$ when pairing a given conditioned stimulus CS1 to the US. In phase 2 ($\la=0$), we pair a second stimulus CS2 to CS1 and the blue curve runs over trials where the associative strength $V^{\rm CS1+CS2}$ of the combined conditioned stimuli \citep{RW72,WR72} decreases to zero. This is standard inhibitory conditioning. Alternatively, one can consider a backward pairing between the US and a previously untrained stimulus CS$-$. Then, we identify the coordinate $V=:\tilde V+1$ in the figure with the association strength $\tilde V$, shifted by 1, of CS$-$. The blue curve then runs from $\tilde V=0$ to $\tilde V=-1$ (perfect inhibition). Trace conditioning follows a similar rule.

The rest of the set (in black) represents experiments where trials with excitatory pairing (single CS followed by the US) are mixed with trials where the US does not follow the CS. The learning curve will change according to the session pattern (positive or negative contingency of the US) and one will have infinitely many possible experiments or natural situations with non-optimal learning. 

By interpreting a model of Pavlovian conditioning as a collection of processes taking place on a fractal set, we gain a number of insights that can be described in a very minimalistic but effective fashion. For instance, the abstract\footnote{In the real world, excitation and inhibition are not symmetrical, since any inhibitory process must depend on preceding excitation. However, in the present case we are interested in a reformulation of the Hull and Rescorla--Wagner \emph{mathematical} models.} notion that excitation and inhibition are the two extremes, with opposite sign, of the same process \citep{Res67} translates into a precise mathematical statement. For the Hull and Rescorla--Wagner models, excitation and extinction correspond to processes living on, respectively, the two complementary branches $\cS_1(I)$ and $\cS_2(I)$ of the Cantor set with initial condition $V_1^{\rm excit}=0$ and $V_1^{\rm extin}=1$.

Also, the Hausdorff dimension of $\cC$ is equal to the capacity, which is found from Eqs.\ \Eq{dc} and \Eq{conpa}:
\be\label{dc2}
\boxd{\dh(\cC)=\dc(\cC)=-\frac{\ln 2}{\ln(1-\a\b)}\,.}
\ee
Another but more rigorous way to find the Hausdorff dimension is via the summation formula \citep{Fal03}
\be\label{sumfo}
1=a_1^{\dh}+a_2^{\dh}= 2(1-\a\b)^{\dh},
\ee
where in the last step we used the similarity ratios \Eq{conpa}. This formula is more general than that of the capacity and can be applied also to the case of more than two similarities acting on a $D$-dimensional space $\mathbb{R}^D$, $\sum_{k=1}^K a_k^{\dh}=1$.

\begin{figure}
\centering
\includegraphics[width=7.8cm]{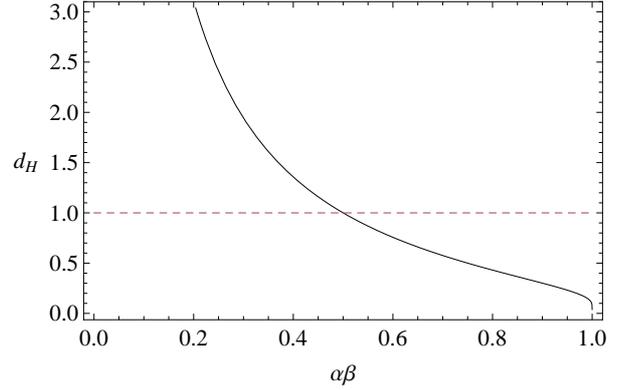}
\caption{\label{fig2} The Hausdorff dimension (or capacity) \Eq{dc2} (solid line), which measures the efficiency of conditioning. The dashed line $\dh=1$ is shown for reference. The region where $\dh>1$ is excluded in fractal geometry but it acquires a meaning in our context.}
\end{figure}

In the example of Fig.\ \ref{fig1}, $\dc=\ln 2/\ln 3\approx 0.63$. The profile \Eq{dc2} is shown in Fig.\ \ref{fig2}.

Note that $\dh$ is independent of the magnitude $\la$ of the US but depends on the learning-rate parameters $\a$ and $\b$, the salience of the CS and of the US. We can identify two regimes:
\begin{itemize}
\item For $0<\a\b<0.5$, the fractal is degenerate, in the sense that it fills the whole line. A lower salience of the CS increases the similarity ratio $a=1-\a\b$ and produces longer segments, leading to a Cantor set with a larger dimension. Strictly speaking, although $\dh>1$, the dimension of the set is just equal to 1, since the continuous line fixes, so to speak, the maximum occupation of points and it cannot be overflown. However, conditioning makes sense in this region and $\dh$ is a good learning index even in this range of values. The latter is consistent both with a CS with very low salience $\a$ (resulting in slow conditioning) and with the expectation that, at the beginning of the training, a very salient CS would actually compete with the US (very small $\b$ versus relatively large $\a$) rather than being a neutral stimulus. For $\b=1$ and the typical CS salience $\a=0.5$, one has $\dh=1$.
\item For $\a\b>0.5$, we have $0<\dh<1$. The closer the value of $\a\b$ to 1, the larger the denominator of Eq.~\Eq{dc2}: as $\a\b\to 1$ and $1-\a\b\to 0$, we move towards the limit $\dh\to 0$. The interpretation of this feature in terms of Pavlovian conditioning is straightforward. The larger the salience of the US, the more efficient the conditioning and the fewer the trials needed to achieve complete training. Perfect training corresponds to infinitely many iterations, but efficient training requires less iterations to reach optimal learning. But the fewer the trials, the ``fewer the points'' in $\cC$, which means the smaller the dimension $\dh$.
\end{itemize}
We have thus established that \emph{the capacity or Hausdorff dimension of the fractal associated with the conditioning model decreases with the increase in the efficiency of the training}. We conjecture that this conclusion may be valid for any other conditioning that can be defined by one or more iterative equations. Of course, different models will correspond to different fractals, not necessarily Cantor dusts.

Interestingly, our result \Eq{dc2} does not depend on the magnitude $\la$ of the US, which is another factor affecting the efficiency of a conditioning. We must conclude that the capacity dimension is insufficient to fully characterize the efficiency, which is a $\la$-dependent concept. However, it is a first step towards a geometric classification of conditioning methods.


\section{Deterministic generalizations}\label{general}

The purpose of this section is to illustrate, without entering into too many details, the potential that the powerful techniques of fractal geometry can express in the context of learning theory and behavioral models. In Section \ref{geoin}, we considered treatments with a single conditioned stimulus associated with the strength parameter $V_n$ governed by the linear equation \Eq{rewam}. We translated both features into two mathematical concepts. (A) Having only one CS allowed us to describe the learning process in terms of a set $\cC$ that can be completely embedded \emph{in one dimension, i.e., a line}. (B) The linearity of Eq.\ \Eq{rewam} led to the \emph{linear} mappings \Eq{contr}; a mapping $\cS(x)$ is linear when it only has the constant and the $x$ term as in \Eq{contr}. Mappings with terms proportional to $x^2,x^3,\dots$, or with entangled variables appearing in mixed terms $xy,xy^2,x^2y,\dots$, are called nonlinear. However, in mathematics and physics there exist many one-dimensional\footnote{Here and in the following, with \emph{one-dimensional} and \emph{multidimensional} we refer to the topological dimension $D$ of the \emph{embedding} space $\mathbb{R}^D$, i.e., the number of independent variables entering the iterative equations defining the fractal. Embedding space is the ambient ``box'' (a line, a plane, a three-dimensional space, and so on) in which the fractal is imagined to live.} fractals that are described by nonlinear (rather than linear) mappings, thus breaking condition (B). Also, the majority of fractals do not fit into a line and they need a higher-dimensional embedding space, thus breaking both (A) and (B). 

Mathematically, nonlinear mappings are the most natural way to generalize the Cantor construction to generic one-dimensional deterministic fractals $\cF$. Just like the Cantor set, deterministic fractals can be defined as the union of the image of several maps, $\cF=\cS_1(\cF)\cup \cS_2(\cF)\cup\cdots\cup\cS_K(\cF)$, but now the functions $\cS_k(x)$, $k=1,\dots,K$, are contractions nonlinear in $x$. A map $\cS$ is called a \emph{contraction} if there exists a constant $0<a<1$ such that $|\cS(y)-\cS(x)|\leq a|y-x|$. When the inequality is saturated (i.e., $|\cS(y)-\cS(x)|= a|y-x|$), we have a similarity, of which the formul\ae\ in \Eq{contr} are an example [in Eq.\ \Eq{comap}, $a=1/3$].
 
Many dynamical system studied in chaos theory are encoded into nonlinear maps. Here, fractals arise in a rather subtle way [see, e.g., Chap.\ 13 of \citet{Fal03}]. Consider, for instance, the function $x_{n+1}=f(x_n)=g x_n(1-x_n)$ on the real line, where $g>0$ is a constant. This is called the logistic map and it was first proposed to model the population growth of certain animal species. A system like this is called chaotic because its behavior is strongly affected by the value of $g$, and even tiny changes in $g$ can lead to very different behaviors. For some values, the dynamics can be highly sensitive to the initial conditions, so that acting with $f$ on two points in the same neighborhood quickly leads, after only a few iterations of $f$, to a very different evolution. In particular, for $g$ greater than some critical value $g_*$ the nonlinear Cantor set $\cC_{\rm nl}= \cS_+(\cC_{\rm nl})\cup\cS_-(\cC_{\rm nl})$ defined by the two mappings $\cS_\pm(x)=1/2\pm\sqrt{1/4-x/g}$ is a chaotic repeller of $f$. A repeller of a dynamical system described by some $f$ is a set $\cF$, invariant under $f$ [this means that $f(\cF)=\cF$], such that points outside $\cF$ are mapped away from it. The nonlinear Cantor set is invariant under the logistic map, that is, $f[\cS_\pm(x)]=x$ for all $x\in[0,1]$; it is not difficult to show that it is a chaotic repeller for $f$. From this example, one can appreciate two things: that $\cC_{\rm nl}$ is defined by mappings $\cS_\pm(x)$ nonlinear in $x$, and that it arises as a special set of points in a dynamical system described by the nonlinear logistic map $f$. 

The logistic map and other one-dimensional nonlinear mappings $f(x)$ used in biology and economics can be found in \citet{Fal03} and in the interpretative review by \citet{May76}, one of the early seminal papers on chaos theory. Examples of multidimensional systems described by mappings mixing coordinates nonlinearly, and where fractals appear as dynamical repellers, are the ``baker's transformation'', Hénon's map, and the solenoid \citep{Fal03}. Multidimensional chaotic systems have applications not only in biology and economics, but also in cryptographic systems \citep{MiMi,RhS}.

The correspondence between learning models and fractal geometry found in Section \ref{geoin} was limited to Hull's case. The question we wish to ask ourselves is: Can we extend it further? Can fractals in chaotic systems (such as those mentioned above) correspond to some learning models in the psychological literature, or are they just mathematical complications with no practical interest? We argue in favor of a positive answer. In this section, we discuss precisely these generalizations of the one-dimensional condition (A) and of the linearity condition (B) to multidimensional models described by many-variables recursive equations (Sections \ref{rwmod1} and \ref{rwmod2}). A simplified learning scenario which is one-dimensional but describable by a nonlinear equation will be presented in Section \ref{nolin}. Our contribution here will be limited to recognize all these learning models as multidimensional and/or nonlinear iterative systems. To show that they are associated to fractals, or even to chaos, is highly nontrivial, but the form of the iterative equations, the explicit results for the nonlinear approximation of Section \ref{nolin}, and other arguments we will advocate below, strongly suggest that fractal geometry waits just beyond the corner. Due to their complexity, we will only sketch future possibilities for multidimensional models, while we will describe the one-dimensional nonlinear generalization in greater detail, showing that it is fractal.


\subsection{Multidimensional systems: Rescorla--Wagner model}\label{rwmod1}

The generalization from fractals on the line to multidimensional fractals captures situations where learning is described by more than one internal variable. Simply put, instead of having only one association strength related to one CS [the one-dimensional condition (A)], we can consider many CSs each with its own association strength (Rescorla--Wagner model) or stimuli with varying salience (Mackintosh model) and magnitude (Pearce--Hall model).

Let us examine first the case of many CS. The Rescorla--Wagner model was proposed to describe the case of compound stimuli, in which case the iterative evolution is more complicated. For two cues A and B with salience $\a_{\rm A}$ and $\a_{\rm B}$, one has two iterative processes $V_n^{\rm (A)}$ and $V_m^{\rm (B)}$, with Eq.\ \Eq{DeV} replaced by
\bs\label{rw2c}\ba
\De V_n^{\rm (A)}=\a_{\rm A}\b[\la-(V_{n-1}^{\rm (A)}+V_{n-1}^{\rm (B)})]\,,\\
\De V_m^{\rm (B)}=\a_{\rm B}\b[\la-(V_{m-1}^{\rm (A)}+V_{m-1}^{\rm (B)})]\,.
\ea\es
Clearly, this extension of the single-CS case is highly nontrivial if A and B are not presented together at all sessions. Also, at different phases one might want to couple different CSs with different USs. The above pair of equations would then be augmented by another identical pair with a different US with salience $\tilde\b$ and asymptote (intensity) $\tilde\la$, and possibly a different CS compound A$\tilde {\rm B}$. Each conditioned stimulus ${\rm CS}^{(i)}$, $i=1,\dots,D$, corresponds to an association strength $V^{(i)}$, which parametrizes the $i$th direction of the $D$-dimensional embedding space wherein the ``Rescorla--Wagner fractal'' lives.

Systems of recurrence equations with $D$ variables $V_{n}^{(i)}$ can be much more difficult to solve analytically than stand-alone expressions such as \Eq{rewam}, depending on how such variables are mutually entangled. Conceptually, there should be no problem in extending our geometric interpretation and one may still be able to construct fractal sets by joining excitatory and inhibitory branches. However, the proof of this involves either the product of $D$ one-dimensional fractals or the study of $D$-dimensional nondecomposable fractals, both of which cases require a machinery more sophisticated than the one developed here for one-dimensional fractals \citep{Fal03}. Nevertheless, we have found a simple result for the case where all CS are presented simultaneously. We begin by observing that the two-cue Rescorla--Wagner model \Eq{rw2c} is solvable analytically in this simplified setting. Consider the phase of an experiment where both cues A and B are presented at the same time and at each trial, cue A starting with $V_1^{\rm (A)}=0$ and cue B with some generic value $0\leq V_1^{\rm (B)}\leq\la$. Then, from Eq.\ \Eq{rw2c} we can exactly solve the system $V_n^{\rm (A)}=V_{n-1}^{\rm (A)}+\De V_n^{\rm (A)}$, $V_n^{\rm (B)}=V_{n-1}^{\rm (B)}+\De V_n^{\rm (B)}$:
\ba
V_n^{\rm (A)}\!\!\!\!\! &=&\!\!\!\!\! \frac{(\la-V_1^{\rm (B)})\a_{\rm A}\{1-[1-(\a_{\rm A}+\a_{\rm B})\b]^{n-1}\}}{\a_{\rm A}+\a_{\rm B}}\,,\nonumber\\
V_n^{\rm (B)}\!\!\!\!\! &=&\!\!\!\!\! \frac{V_1^{\rm (B)}\a_{\rm A}+\a_{\rm B}\{\la-(\la-V_1^{\rm (B)})[1-(\a_{\rm A}+\a_{\rm B})\b]^{n-1}\}}{\a_{\rm A}+\a_{\rm B}}\,.\nonumber
\ea
Each of these two solutions can combine separately into the linear recursive equations
\bs\label{rw2c2}\ba
\hspace{-.7cm}V_n^{\rm (A)}\!\!\!\!\! &=&\!\!\!\!\! [1-(\a_{\rm A}+\a_{\rm B})\b]V_{n-1}^{\rm (A)}+(\la-V_1^{\rm (B)})\a_{\rm A}\b\,,\label{rw2c2a}\\
\hspace{-.7cm}V_n^{\rm (B)}\!\!\!\!\! &=&\!\!\!\!\! [1-(\a_{\rm A}+\a_{\rm B})\b]V_{n-1}^{\rm (B)}+(\la\a_{\rm B}+V_1^{\rm (B)}\a_{\rm A})\b.\label{rw2c2b}
\ea\es
Therefore, the system \Eq{rw2c} with coupled variables $V_n^{\rm (A)}$ and $V_n^{\rm (B)}$ has been recast as the pair \Eq{rw2c2} of \emph{independent} similarity maps. Coupling Eq.\ \Eq{rw2c2a} with its extinction counterpart and recalling Eq.\ \Eq{dc2}, we obtain a Cantor set $\cC_{\rm A}$ with dimension $\dh(\cC_{\rm A})=-\ln 2/\ln[1-(\a_{\rm A}+\a_{\rm B})\b]$. Doing the same with Eq.\ \Eq{rw2c2b}, we get another copy $\cC_{\rm B}$ of the same set with dimension $\dh(\cC_{\rm B})=\dh(\cC_{\rm A})$. Recalling that the Hausdorff dimension of the product of two Cantor sets is the sum of their dimensions \citep{Fal03}, we conclude that the two-cue Rescorla--Wagner model is associated with a set $\cC_{\rm A}\times \cC_{\rm B}$ with dimension
\be
\dh(\cC_{\rm A}\times \cC_{\rm B})=\dh(\cC_{\rm A})+\dh(\cC_{\rm B})=-\frac{2\ln 2}{\ln[1-(\a_{\rm A}+\a_{\rm B})\b]}.\nonumber
\ee
Now, notice that the inclusion of an arbitrary number $D$ of cues would not change the above argument: the only change would be in the initial conditions $V_1^{(i)}$, which would affect the constants $b_k$ in the similarity maps $\cS_k(x)=a_k x+b_k$. Then, it is easy to see that the multi-cue Rescorla--Wagner model with all CS presented at each trial is associated with the set $\cC_{\rm RW}=\prod_{i=1}^D \cC_i$ (the product of $D$ Cantor sets) with Hausdorff dimension
\be
\boxd{\dh(\cC_{\rm RW})=\sum_{i=1}^D\dh(\cC_i)=-\frac{D\ln 2}{\ln\left[1-\left(\sum_{i=1}^D\a_i\right)\b\right]}\,.}
\ee
This formula is valid only if $\sum_i\a_i<1/\b$. For instance, for a US with $\b=1$ and two cues, it must be $\a_{\rm A}+\a_{\rm B}<1$. 

For experimental designs with nonsimultaneous presentation of all cues at all trials, with cues with too large salience, or with different USs, this simplified model breaks down and the analysis can become considerably difficult. In general, the Hausdorff dimension of the product of many sets cannot be determined exactly and the only thing one can do is to bound it from above and below by certain combinations of the dimensions of the sets \citep{Fal03}.


\subsection{Multidimensional systems: Mackintosh and Pearce--Hall models}\label{rwmod2}

There are other examples of multidimensional systems. It is well-known that the Rescorla--Wagner model suffers from several limitations \citep{MBG}. Among them, we recall that it predicts the extinction of conditioned inhibition (which does not occur actually) and it regards extinction as an unlearning process (i.e., the blue and red curves in Fig.\ \ref{fig1} are perfectly specular). Thus, it cannot explain either spontaneous recovery (when a CR that had been extinguished reappears) or other effects such as preconditioning exposure to the CS [i.e., latent inhibition, which may occur also in conjunction with a reinforcer \citep{HP}], augmentation (or counter-blocking), first-trial unblocking \citep{Ma75a}, or unblocking by the surprising omission of part of a compound US \citep{DHM}. Moreover, the blocking effect on a CS2 (by a CS1 associated with the US in a preliminary training phase) is explained as the absence of novelty in the US after its pairing with the CS1, but this interpretation has been ruled out in an experiment by Mackintosh and Turner \citeyearpar{MacT}. Latent inhibition was accounted for by \citet{Wag78}, while first-trial unblocking and unblocking by omission were explained by Mackintosh model of attention \citep{Mac75}, according to which blocking occurs because predictive but redundant stimuli such as CS2 are ignored. Finally, Pearce and Hall \citeyearpar{PH} proposed a model that could encompass all these cases and explain latent inhibition as well as various phenomena of unblocking. These and other elemental theories of associative learning are reviewed by \citet{LeP} and Wagner and Vogel \citeyearpar{WV}; some have been developed more recently \citep{EH}.

Most of these proposals require a quantitative modification of the Rescorla--Wagner model by replacing all constant $\a$'s with trial-dependent parameters $\a_n$ that change with the subject's experience. In the case of Mackintosh model, the rate of change is assumed to follow a linear law. For instance, given two cues A and B, one has
\be\label{compAB}
\De\a^{\rm (A)}_n=\g_{\rm A}\left(\left|\la-V_{n-1}^{\rm (B)}\right|-\left|\la-V_{n-1}^{\rm (A)}\right|\right)
\ee
and an analogous expression for $\De\a_n^{\rm (B)}$, where $\g_{\rm A}$ is a constant. The recursive law governing the evolution of $V^{\rm (A)}$ is $\De V_n^{\rm (A)}=\a_{n-1}^{\rm (A)}\b[\la-V_{n-1}^{\rm (A)}]$ and the influence of other stimuli is encoded exclusively in the way $\a^{\rm (A)}$ varies, via Eq.\ \Eq{compAB}. This is in contrast with the Rescorla--Wagner prescription, according to which the size of the associative change depends on the strength of all the stimuli, $\De V_n^{\rm (A)}=\a_{\rm A}\b[\la-\sum_iV_{n-1}^{(i)}]$.

According to Mackintosh model, attention is competitive and is based on relative predictivity of different cues. For illustrative purposes only, here we are interested in a simpler model with less variables (i.e., dimensions $D$ of the dynamical system; in the two-cue model \Eq{compAB}, $D=3$). The outcome will be unrealistic for several reasons, but we stress that the following simplified model should be regarded as a sketch of the possibility that the multidimensional iterative systems employed in conditioning psychology can admit a fractal reinterpretation. Moreover, this simplified ``Mackintosh'' model (quotation marks are due) will prepare the ground to a more interesting generalization, namely, to nonlinear systems.

\begin{figure}
\centering
\includegraphics[width=7.8cm]{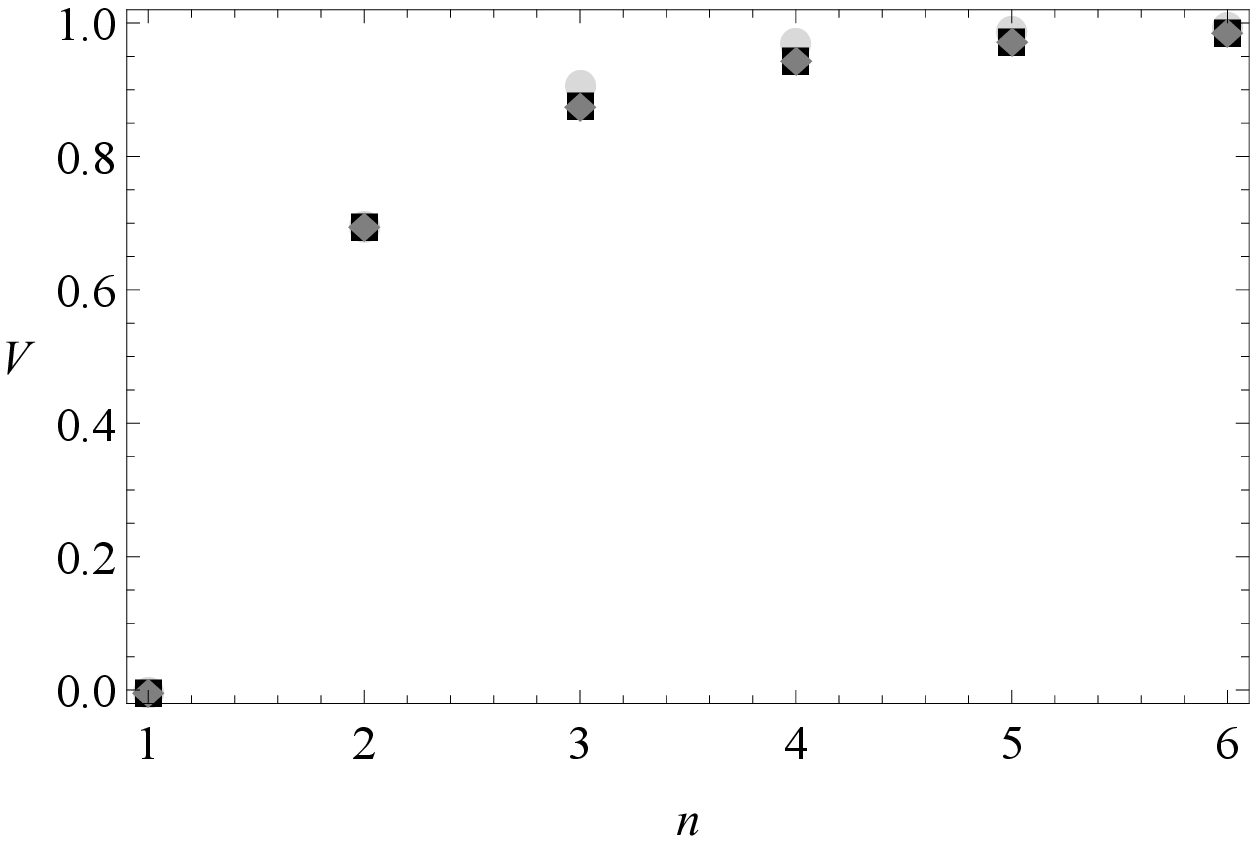}
\caption{\label{fig3} Learning progression for excitatory conditioning with $V_1=0$, $\b=1$ and $\la=1$. Light-gray circles: Hull's model \Eq{lacu1} for $\a=0.7$. Black squares: ``Mackintosh'' single-cue model \Eq{rewam} and \Eq{rewama} for $\a_1=0.7$ and $\g=0.099$ (the asymptote is $\a_n\to 0.5503$). Gray diamonds: the nonlinear model \Eq{rewam2} with $\tilde\la=\la$, $\a_{\rm min}=0.5503$, and $\a_{\rm max}= 0.7$.}
\end{figure}

With this disclaimer in mind, if we ignore stimulus B we get a $D=2$ single-cue case, which is better tractable:
\be\label{rewama}
\a_n=\a_{n-1}+\De \a_n=\a_{n-1}-\g|\la-V_{n-1}|\,,
\ee
which should be coupled with Eq.\ \Eq{rewam}. Here $\g>0$ is a constant. This system has a two-dimensional embedding where the $D=2$ directions are the association strength $V$ and the CS salience $\a$. Although there is no cue competition and we cannot thus apply the actual Mackintosh model, the other main tenet of Mackintosh is implemented, namely, that $\a$ varies with the subject's experience and depends on the correlation of the CS (the only cue in a phase of an experiment) with the reinforcement \citep{Mac75}. In this particular case, $\a$ decreases while the associative strength increases towards the asymptote, contrary to Mackintosh two-cue model where $\a$ typically increases for a cue which is a better predictor of the outcome than are all other presented cues. We were unable to find explicit solutions $V_n$ and $\a_n$ but it is easy to see that, compared with the Hull learning curve \Eq{lacu1} with $\a=\a_1$, this model predicts a slower learning during intermediate trials (Fig.\ \ref{fig3}). 

The model \Eq{rewama} is a hybrid between Mackintosh and Pearce--Hall. Also the Pearce--Hall model assumes that the effectiveness $\a$ of a CS changes with its predictive strength $V$ but, contrary to Mackintosh theory, the magnitude or intensity of the US now varies with the experience \citep{PH}: $\a_n=|\la_{n-1}-V_{n-1}|$, where the US magnitude is bound to lie in the range $0\leq\la_n\leq 1$. In a variant of the model which fixes some issues of the original proposal, this expression is replaced by $\a_n=\g|\la_{n-1}-V_{n-1}|+(1-\g)\a_{n-1}$, where $0<\g\leq 1$ \citep{PKH}. This reflects the idea that stimuli always have the possibility to gain access to the subject's processing, but less surprising stimuli will have limited access. The recursive law for $\a_n$ is combined with $\De V_n=\b\a_n\la_n$, to give a three-dimensional system ($D=3$) with three directions parametrized by $V$, $\a$, and $\la$.
 Without any specific iteration rule for $\la_n$ [absent in the original paper by Pearce and Hall \citeyearpar{PH}], we cannot solve the system analytically. However, when $\la_n$ is approximately constant we get a single-cue setting with decreasing $\a$, just like the model \Eq{rewama}.

Having observed that the Rescorla--Wagner, Mackintosh, and Pearce--Hall models are multidimensional iterative systems, it remains to see that they correspond to fractals. As said in the introduction of this section, this check lies out of the range of the present investigation, which is exploratory in nature. However, it is very likely that an underlying fractal geometry exists, for three reasons. First, from a visual inspection of the above equations, but which could be revealing only for a mathematician. Second, because the simplified core of all these models, Hull's model, is already a neat example of fractal. Third, the model presented in Section \ref{nolin} approximates one of the multidimensional models (Mackintosh's) and is manifestly associated with a fractal. The point is that generalizing a one-dimensional model to many dynamical variables and/or to nonlinear mappings is not expected to change (and it does not, as shown below) the geometry from fractal to smooth.


\subsection{One-dimensional nonlinear systems: a nonlinear learning model}\label{nolin}

A somewhat easier but still nontrivial possibility is to analyze learning processes which are one-dimensional but described by nonlinear recurrence equations. The following example can help the reader to appreciate that such abstract constructions can have direct applications to psychology. We present (to the best of our knowledge, for the first time) a simplified model of variable salience which approximates single-cue ``Mackintosh'' model. Equation \Eq{rewama} is replaced by
\be\label{alphan}
\a_n=\a_{\rm max}-(\a_{\rm max}-\a_{\rm min})\frac{V_n}{\la}\,,\quad  0<\a_{\rm min}<\a_{\rm max}\leq 1.
\ee
The parameter $\a_n$ depends on the value $V_n$ of the association strength at the moment of the trial. Using Eq.\ \Eq{DeV}, it is easy to see that \Eq{alphan} is an approximation of \Eq{rewama} when $\a_n$ does not vary much during conditioning ($\a_{\rm max}\sim\a_{\rm min}$), corresponding to a small parameter
\be\nonumber
\g\simeq(\a_{\rm max}-\a_{\rm min})\frac{\b}{\la}\ll 1\,.
\ee
In excitatory conditioning, when $n=1$ (first trial), $V_1=0$ and the salience of the CS has some default value $0<\a_1=\a_{\rm max}<1$ which depends on the nature of the CS and on its salience for the subject. As the excitatory training proceeds, the salience of the CS approaches the asymptotic value $\a_{\rm min}<\a_1$. In the case of extinction, the association strength decreases from $V_1=\la$ to zero and the salience of the CS grows from its minimal value $\a_1=\a_{\rm min}$ up to $\a_{\rm max}$. Therefore, if the subject is presented with a novel stimulus, the salience will decrease from $\alpha_{\rm max}$ to $\alpha_{\rm min}$ ($V$ increasing from 0 to $\la$), while in the case of extinction ($V$ decreasing from $\la$ to 0) the converse will happen.\footnote{For inhibitory conditioning, either one makes the change of variables $V=\tilde V+1$ explained below Eq.\ \Eq{conpa} or one considers the compound case \Eq{compAB}.} Plugging Eq.~\Eq{alphan} into \Eq{rewam}, we obtain a nonlinear law for excitatory conditioning:
\ba
V_n&=&(1-\a_{n-1}\b)V_{n-1}+\a_{n-1}\b\la\nonumber\\
&=&\a_{\rm max}\b\la+[1+\b(\a_{\rm min}-2\a_{\rm max})]V_{n-1}\nonumber\\
&&+(\a_{\rm max}-\a_{\rm min})\frac{\b}{\la}V_{n-1}^2.\label{rewam2}
\ea
For small $\g$, this is a good approximation of single-cue Mackintosh model (Fig.\ \ref{fig3}). To resume, the nonlinear model \Eq{rewam2} reduces to the Hull/Rescorla--Wagner model only at lowest order in the approximation, when $\a_n=\a$ is exactly constant. However, in Eq.\ \Eq{alphan} $\a_n$ has a linear dependence on $V$, which translates into the nonlinear term $O(V^2)$ in the evolution equation \Eq{rewam2} for the association strength. Figure \ref{fig3} clearly shows that the nonlinear model is, on one hand, a very good approximation of single-cue Mackintosh's for a slowly varying $\a$ (small parameter $\g$; gray diamonds overlap completely with black squares) and, on the other hand, distinctly different with respect to Hull model (light-gray circles). The conditioning described by Eq.~\Eq{rewam2} differs from other nonlinear models described in the past \citep{BVW,LeP}. In particular, Le Pelley's hybrid model \citep{LeP} is an extension, rather than an approximation, of Mackintosh theory.

The excitation branch $\tilde\cS_1(I)$ of the associated fractal (in this case, a nonlinear Cantor set we will dub $\tilde\cC$) is given by the contraction \Eq{rewam2}, while the extinction branch $\tilde\cS_2(I)$ is given by setting $\la=0$ into Eq.\ \Eq{rewam} and then plugging \Eq{alphan} therein:
\ba
\hspace{-.8cm} V_n&=&(1-\a_{n-1}\b)V_{n-1}\nonumber\\
\hspace{-.8cm} &=&(1-\b\a_{\rm max})V_{n-1}+(\a_{\rm max}-\a_{\rm min})\frac{\b}{\la}V_{n-1}^2\,.\label{rewam3}
\ea
Then, the set $\tilde\cC=\tilde\cS_1(\tilde\cC)\cup\tilde\cS_2(\tilde\cC)$ is given by the two mappings
\bs\label{newsys}\ba
\hspace{-.8cm}\tilde\cS_1(x)&=& \a_{\rm max}\b\la+\left[1+\b(\a_{\rm min}-2\a_{\rm max})\right]x\nonumber\\
\hspace{-.8cm}&&+(\a_{\rm max}-\a_{\rm min})\frac{\b}{\la}x^2\,,\\
\hspace{-.8cm}\tilde\cS_2(x)&=& (1-\b\a_{\rm max})x+(\a_{\rm max}-\a_{\rm min})\frac{\b}{\la}x^2\,.
\ea\es
When $\a_{\rm max}=\a_{\rm min}$, we recover $\cS_1$ and $\cS_2$. Notice that, in general, the appearance of two different conditioning laws for excitatory and inhibitory processes is nothing new and it was already employed in Pearce and Hall \citeyearpar{PH}. The novelty here, apart from the specific form of Eq.~\Eq{newsys}, is the geometric interpretation of conditioning in terms of branches of a fractal. As an application, we now show that the capacity of this set is larger than the one of Hull's model, thus giving a quantitative estimate of the ``uphill learning'' depicted in Fig.\ \ref{fig3}.\footnote{The reason behind this name and the details of the approximation linking Mackintosh's model and Eq.\ \Eq{rewam} will be discussed in a separate publication.} Although it is not always possible to find the exact value of the dimension of a fractal, there are some powerful theorems that make use of the properties of contractions. The reader uninterested in technicalities can skip this part and go directly to Eq.\ \Eq{dimes}. 

The first step consists in checking whether the maps $\tilde\cS_k$ are bi-Lipschitz, meaning that there exist two positive and finite constants $a_k$ and $b_k$ such that $b_k|y-x|\leq |\tilde\cS_k(y)-\tilde\cS_k(x)|\leq a_k|y-x|$. In that case, $a_k={\rm sup}_{x}|\tilde\cS_k'(x)|$ and $b_k={\rm inf}_{x}|\tilde\cS_k'(x)|$, where a prime denotes the derivative with respect to $x$. For the system \Eq{newsys}, it is easy to find these constants. Since $\tilde\cS_1'(x)=1+\b(\a_{\rm min}-2\a_{\rm max})+2(\a_{\rm max}-\a_{\rm min})({\b}/{\la})x$, $\tilde\cS_2'(x)=1-\b\a_{\rm max}+2(\a_{\rm max}-\a_{\rm min})({\b}/{\la})x$, $\a_{\rm max}-\a_{\rm min}>1$, and $0\leq x\leq1$, one has that the highest and lowest (respectively sup and inf) value of $|\tilde\cS_k'(x)|$ is attained at, respectively, $x=1$ and $x=0$. Then, we find $a_1=|1+\b(\a_{\rm min}-2\a_{\rm max})+2\b(\a_{\rm max}-\a_{\rm min})/\la|$, $a_2=|1-\b\a_{\rm max}+2\b(\a_{\rm max}-\a_{\rm min})/\la|$, $b_1=|1-\b(2\a_{\rm max}-\a_{\rm min})|$, and $b_2=1-\b\a_{\rm max}$. For $\la=1$ and assuming that there is not much difference between the initial and final value of $\a_n$, these expressions reduce to
\bs\label{a1b1}\ba
\hspace{-1cm}&&a_1=1-\b\a_{\rm min}\,,\quad a_2=1+\b(\a_{\rm max}-2\a_{\rm min}),\\
\hspace{-1cm}&&b_1=1-\b(2\a_{\rm max}-\a_{\rm min})\,,\quad b_2=1-\b\a_{\rm max}.
\ea\es
In particular, for Hull's model $\a_{\rm min}=\a_{\rm max}=\a$ and all the coefficients collapse to $1-\b\a$. Next, we prove that the nonlinear model is associated with a fractal. This check is important because, if there is an underlying fractal, then by approximation also Mackintosh model corresponds to a fractal geometry, which yields support to the main claim of this section. To show this, one must calculate the Hausdorff dimension and find a noninteger value. One recalls that the Hausdorff dimension of a fractal $\cF=\cS_1(\cF)\cup\cS_2(\cF)$ is bounded from above and from below by $s_b\leq\dh(\cF)\leq s_a$, 
 where $s_b$ and $s_a$ are two constants determined implicitly by the relations [analogous to \Eq{sumfo}] $b_1^{s_b}+b_2^{s_b}=1$ and $a_1^{s_a}+a_2^{s_a}=1$ \citep{Fal03}. For linear mappings, $s_a=s_b$ and these relations are sufficient to determine $\dh$. For nonlinear mappings, one can at least make an estimate of the range of $\dh$. For instance, consider the bi-Lipschitz maps \Eq{newsys} with $\b=1$, $\la=1$, $\a_{\rm min}=0.55$ and $\a_{\rm max}=0.70$ in Eq.\ \Eq{a1b1}. Then, the parameters \Eq{a1b1} are fully determined and one finds $0.455<\dh(\tilde\cC)<1.077$. Taking an extra iteration and the maps $\{\tilde\cS_k\circ\tilde\cS_l:k,l=1,2\}$, this interval is refined to $0.508<\dh(\tilde\cC)<0.932$, while a third iteration with the maps $\{\tilde\cS_k\circ\tilde\cS_l\circ\tilde\cS_q:k,l,q=1,2\}$ yields $0.546<\dh(\tilde\cC)<0.857$. A fourth iteration with $\{\tilde\cS_k\circ\tilde\cS_l\circ\tilde\cS_q\circ\tilde\cS_r:k,l,q,r=1,2\}$ gives $0.572<\dh(\tilde\cC)<0.814$. It is not difficult to convince oneself that
$0.576<\dh(\tilde\cC)<0.868$, where inequalities are strict and the lower and upper limit correspond to the Hausdorff dimension \Eq{dc2} of the ternary Cantor set with, respectively, $\a=0.70$ and $\a=0.55$. For general CS saliences $\a_{\rm min}$ and $\a_{\rm max}$ and a fixed US salience $\b$,
\be\label{dimes}
\boxd{\dh(\cC_{\a=\a_{\rm max}})<\dh(\tilde\cC)<\dh(\cC_{\a=\a_{\rm min}})\,.}
\ee
Numerical iterative methods such as that above can shrink this range considerably. Therefore, the nonlinear model describes less efficient learning than Hull's when its US salience is smaller than that of Hull's models, and vice versa. This conclusion is obvious by looking at Fig.\ \ref{fig3} but we have just made it quantitative in a precise sense. The main point, however, is that Eq.\ \Eq{dimes}, which is a nontrivial consequence of the theorem cited above, proves that the dimension of $\tilde\cC$ is noninteger. Since this set is defined by the action of two contractions, one concludes that $\tilde\cC$ is a deterministic fractal.


\section{Random fractals}\label{rafra}


\subsection{Random Cantor sets: varying programs}\label{rafra1}

It is easy to generalize the construction of Sections \ref{geoin} and \ref{general} to other experiments. For instance, an asymmetric Cantor set is obtained not only in the nonlinear model proposed above, but also in the linear case if we choose different saliences $\a_2\neq \a_1$ and $\b_2\neq\b_1$ in the parameters \Eq{conpa} of Eq.\ \Eq{contr}. Also, if we let any of the parameters $\la$, $\a$, and $\b$ vary randomly in the interval $[0,1]$ at each iteration, we would be in a situation where the strength and appearance rate of the US is governed by a random generator at each conditioning trial. Therefore, at each iteration $n$ there are four similarities (one pair $\cS_{1,2}$ per interval),
\bs\label{genran}\ba
&&\cS_{1,{\rm L},n}(x)= a_{1,{\rm L},n} x+b_{1,{\rm L},n}\,,\nonumber\\
&&\cS_{2,{\rm L},n}(x)=a_{2,{\rm L},n} x+b_{2,{\rm L},n}\,,\\
&&\cS_{1,{\rm R},n}(x)= a_{1,{\rm R},n} x+b_{1,{\rm R},n}\,,\nonumber\\
&&\cS_{2,{\rm R},n}(x)=a_{2,{\rm R},n} x+b_{2,{\rm R},n}\,,
\ea\es
the first pair acting on the left-hand interval (L) and the second pair acting on the right-hand interval (R). In the case of Pavlovian conditioning [Eq.~\Eq{conpa}], $a_{1,{\rm L},n}=1-\a_{1,{\rm L},n}\b_{1,{\rm L},n}$, $b_{1,{\rm L},n}=\la_{1,{\rm L},n}\a_{1,{\rm L},n}\b_{1,{\rm L},n}$, and so on.

In the language of fractal geometry, this would be a \emph{random fractal}. We expect the conditioning-to-fractal correspondence to hold only by considering both the excitatory and inhibitory branches at the same time (otherwise, the iterative process would collapse the initial set $I$ to a point). While the ideal excitatory conditioning in a controlled environment is only a portion of the sequence of iterations generating the fractal [in the example of Fig.\ \ref{fig1}, from the interval $(2/3,1)$ to the point $V=1$], we can interpret the whole fractal as a description of the most varied pairings one can find in Nature or in the laboratory between two given stimuli.

There are various ways to randomize a one-dimensional Cantor set. Here we discuss two.
\begin{itemize}
\item One is to divide each interval, starting as usual from $I=[0,1]$, into three equal parts and remove some randomly chosen (even none or all) \citep{Fal03}. Then, for each branch (left L or right R) and at each iteration $n$, there are eight options (Table \ref{tab1}): no subinterval removed ($a_1=a_2=1$, $b_1=b_2=0$), all intervals removed ($a_1=a_2=b_1=b_2=0$), only the central interval removed ($a_1=a_2=1/3$, $b_1=2/3$, $b_2=0$), only the leftmost interval removed ($a_1=b_1=0$, $a_2=2/3$, $b_2=1/3$), only the rightmost interval removed ($a_2=b_2=0$, $a_1=2/3$, $b_1=1/3$), only the leftmost interval surviving ($a_1=b_1=b_2=0$, $a_2=1/3$), only the central interval surviving ($a_1=b_1=0$, $a_2=1/3$, $b_2=1/3$), and only the rightmost interval surviving ($a_2=b_2=b_1=0$, $a_1=1/3$).

In the context of one-dimensional conditioning with constant saliences (one CS, Hull model), this type of randomization is limited by the fact that the coefficients $a_{n}$ and $b_{n}$ are not completely independent (from now on, we keep only iteration indices). $\b\neq0$ is fixed \emph{a priori} (US salience predetermined by the type of stimulus and subject) and we can change $\a_n$ only to the values 0 (absence of CS) or \emph{one} among the three possibilities $\a_{n}=1/\b,1/(3\b),2/(3\b)$ ($a_n=1$ or $2/3$ or $1/3$, presence of CS with a given salience; the case $a_n=1$ is allowed only if $\b=1$, which we can grant). This is because a CS with different saliences (e.g., a light of different colors) is to be treated as many different CSs. The randomizing algorithm can only pick values $\la_{n}$ accommodating with the value of $b_n$, i.e., either 0 (no US) or 1 (maximum magnitude or intensity). Only when no CS is presented ($a_n=1$) can $\la$ get any value between 0 and 1. All these cases are summarized in Table \ref{tab1}.
\begin{table*}
\begin{center}
\begin{tabular}{|l|cccc|cccc|}\hline
Iteration of an interval  											&  $a_1$	 &  $b_1$   & $\a_1\b_1$ & $\la_1$ &  $a_2$	  &  $b_2$   & $\a_2\b_2$ & $\la_2$ \\\hline\hline
\includegraphics[width=4.5cm]{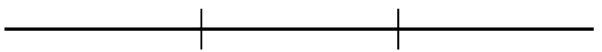}							& 1		  	 & 0				& 0					 & any     & 1		  	& 0				 & 0					& any \\
																								& 0			   & 0				& 1					 & 0	     & 0		  	& 0			 	 & 1					& 0	  \\
\includegraphics[width=4.5cm]{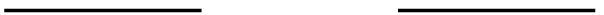}							&$\tfrac13$&$\tfrac23$& $\tfrac23$ & 1	     &$\tfrac13$& 0				 & $\tfrac23$ & 0	  \\
\hspace{1.5cm}\includegraphics[width=3cm]{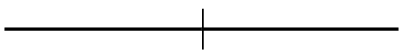}& 0		     & 0			  & 1					 & 0	     &$\tfrac23$&$\tfrac13$& $\tfrac13$ & 1	  \\
\includegraphics[width=3cm]{tab45}						  &$\tfrac23$& 0        & $\tfrac13$ & 0			 & 0  			& 0				 & 1					& 0	  \\
\includegraphics[width=1.5cm]{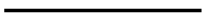}						& 0  			 & 0			  & 1				   & 0       &$\tfrac13$& 0	       & $\tfrac23$ & 0   \\
\hspace{1.5cm}\includegraphics[width=1.5cm]{tab678}& 0	  		 & 0			  & 1				   & 0       &$\tfrac13$&$\tfrac13$& $\tfrac23$ &$\tfrac12$\\
\hspace{3cm}\includegraphics[width=1.5cm]{tab678}&$\tfrac13$&$\tfrac23$& $\tfrac23$ & 1	     & 0			  & 0			   & 1					& 0   \\\hline
\end{tabular}
\caption{\label{tab1}Allowed values of the parameters of random Hull's model in the case of partial reinforcement (random fractal).}
\end{center}
\end{table*}
Clearly, the only combination that makes sense in a single-cue psychological experiment is the first and third line of the table (all subintervals or first and third subinterval present). All the other cases but one are excluded because $\a\b$ is different in the mappings $\cS_1$ and $\cS_2$. The all-or-no-subinterval case (first and second line of the table) is ruled out because it represents no conditioning at all. A set $C_{\rm pr1}$ resulting from the only surviving procedure, together with its excitatory and extinction learning curves, are shown in Fig.\ \ref{fig4}. This corresponds to a controlled experimental design with \emph{partial reinforcement} on a randomized schedule (hence the subscript ``pr'' in $C_{\rm pr1}$), where the CS is either absent or present and the US is either absent or present (with the same intensity except in trials where the CS is absent).
\begin{figure}
\centering
\includegraphics[width=7.8cm]{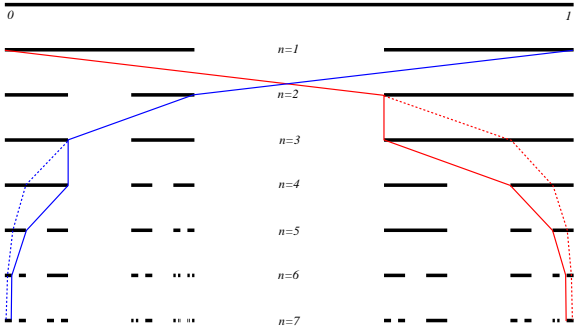}
\caption{\label{fig4} A random ternary Cantor set $C_{\rm pr1}$ corresponding to a randomized Hull model of the first type for $n=7$ iterations. The thin red (blue) solid curve is the learning curve of excitatory conditioning (respectively, extinction) in a partial reinforcement program. Dashed curves are the learning curves in the deterministic model. (For interpretation of the references to color in this figure legend, the reader is referred to the web version of this article.)}
\end{figure}

The Hausdorff dimension of a random Cantor set is given almost surely (i.e., with probability 1) by the expectation value of Eq.\ \Eq{sumfo} $\langle a_1^{\dh}+a_2^{\dh}\rangle=1$ \citep{Fal03}. In the deterministic case, angular brackets are removed and Eqs.\ \Eq{dc2} and \Eq{sumfo} are recovered.  For the similarity ratios \Eq{conpa}, we get $2\langle (1-\a\b)^{\dh}\rangle=1$ and the efficiency of conditioning is increased by decreasing the Hausdorff dimension, that is, by increasing $\a\b$ towards 1 in average. As we already commented, $\dh$ is $\la$-independent and does not capture all aspects of efficiency. For instance, in a program with partial reinforcement we can get the same $\dh$ as in a determinist program with continuous reinforcement (US presented at all trials), but it is known that performance is lower in the first. This phenomenon is clearly shown in Fig.\ \ref{fig4}: the less fragmented is the fractal, the slower the learning. Thus, examination of the detailed properties of the fractal can say more than what said by the global indicator $\dh$.
\item Another possibility is to replace each interval $I_n$ at the $n$th iteration with two subintervals $I_{{\rm L},n+1}$ and $I_{{\rm R},n+1}$ of random length, such that the length ratios $|I_{{\rm L},n+1}|/|I_n|$ and $|I_{{\rm R},n+1}|/|I_n|$ have independent and identical probability distributions for each $n$. If $I_{{\rm L},n+1}$ and $I_{{\rm R},n+1}$ abut, respectively, the left- and the right-hand of $I_n$, then one obtains the set like the one depicted in Fig.\ \ref{fig5}, but this condition is optional. This corresponds to the most general situation such that at each trial the CS may or may not change its salience (i.e., different CS may be presented), the US may or may not be presented, and the US intensity and salience may vary with each presentation. This could be a natural situation where the animal is surrounded by several dynamically evolving stimuli in the environment. Tailoring the algorithm, one can reduce the system to a controlled experimental design (including with partial reinforcement on a randomized schedule) more flexible that the one of the previous case (i.e., we can present stimuli with other values of the saliences and a US with different magnitudes). An example of this randomized Cantor set $C_{\rm pr2}$ is given by Fig.\ \ref{fig5}.
\begin{figure}
\centering
\includegraphics[width=7.8cm]{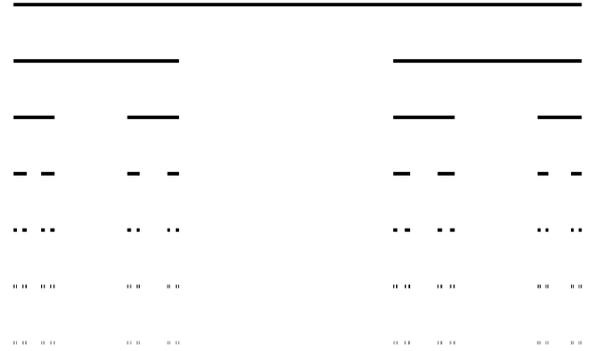}
\caption{\label{fig5} A random Cantor set $C_{\rm pr2}$ corresponding to a randomized Hull model of the second type for $n=6$ iterations and uniform probability distribution.}
\end{figure}
\end{itemize}

There are two factors that may complicate a fractal-based analysis of conditioning processes with a random component: the fact that $\b$ and $\la$ are not independent and the extension to a multi-cue setting. Both factors should be taken into account when planning a realistic simulation, but they do not mar the core of the theory. Concerning the first, we have considered a case where the US is presented on a random schedule ($\la=1$ or $\la=0$ depending on the trial) but $\b$ is fixed. Since $\b$ and $\la$ are both properties of the US, $\b$ could or should vary depending on the presence or absence of the stimulus and, in general, one expects that $\b(\la=1)>\b(\la=0)$ \citep{RW72}, i.e., the US is more salient when is present. Taking this into account can modify the simple partial-reinforcement model we presented above but not qualitatively, as we will see later.

In the presence of more than one CS (Rescorla--Wagner model), trial-varying CS salience (Mackintosh model) and US magnitude (Pearce--Hall model\footnote{Due to an incomplete definition of the Pearce--Hall model, we have not managed to analyze it with our formalism, but we do not expect to find any conceptual difficulty in that direction.}), the random-fractal construction becomes multidimensional and much more varied. For each experimental design, one can draw a unique random fractal on $\mathbb{R}^2$. Its Hausdorff dimension can be estimated as usual but with a given probability $1-q<1$, where $q$ is the probability that the set be empty \citep{Fal03}.


\subsection{Relation with performance variability}\label{relp}

Random fractals are a not-so-old acquaintance in biological sciences and they can be found in physiology (Bassingthwaighte, Liebovitch, \& West, \citeyear{BLW}; Eke, Herman, Kocsis, \& Kozak, \citeyear{EHKK}), neuroscience \citep{Wer10}, and animal behavior and cognitive sciences \citep{Kel10}. Concerning the latter, we wish to comment on whether our formalism may have some application to, or connection with, extant psychological literature on the subject. To see this carefully and in order to avoid hasty conclusions based on apparent but false analogies, we need to make a digression in cognitive science. The aim of comparing our random-fractal learning models with the theory and results of the so-called $1/f$ literature is not just the acknowledgment, for the sake of the record, that fractals have already played a role in psychology. Rather, both in random-fractal learning models and in cognitive performance models variability in the subject response is the main characteristics to be studied in experiments. Understanding what has been done and found in the cognitive field will give us some orientation about possible testable predictions of the fractal paradigm.

The variability (or ``noise'') in human performance in memory tasks, reaction tasks, mental rotation, word naming, and so on, has become a hot trend in recent years [see \citet{RiHo} for a review]. The response of the subject over the sequence of trials can be decomposed by spectral analysis into periodic components with frequency $f$ \citep{EHKK,Hol05,TG}. Experimentally, it was found that the amplitude $\cA$ of these components scale with the frequency according to the power law $|\cA|^2\propto P(f)\sim f^{-\de}$ (Gilden, \citeyear{Gil01}; Gilden, Thornton, \& Mallon, \citeyear{GiTM}; Holden, \citeyear{Hol13}; Holden, Van Orden, \& Turvey, \citeyear{HVOT}; Kello, Anderson, Holden, \& Van Orden, \citeyear{KAHVO}; Kello, Beltz, Holden, \& Van Orden, \citeyear{KBHVO}; Van Orden, Holden, \& Turvey, \citeyear{VOHT1,VOHT2}) [an earlier study highlighting a connection between cognition and fractal geometry is by Kumar, Zhou, and Glaser \citeyearpar{KZG}], where $\de\geq 0$ is a constant (denoted by $\a$ or $\b$ in the literature, all symbols we do not use here to avoid confusion with saliences). The source of this phenomenon (called $1/f$ noise or $1/f$ scaling because in early papers $\de$ was found to be close to 1) is still under debate. At first, it was interpreted as the intrinsic uncertainty, possibly due to an estimation error, in the formation of representations in the mind, such as the reproduction of spatial or temporal intervals in human memory \citep{GiTM}. Different cognitive systems may have different types of idiosyncratic uncertainty, all combining to give $1/f$ noise accidentally; in this interpretation, $1/f$ noise is not a general, fundamental property of human behavior (Farrell, Wagenmakers, \& Ratcliff, \citeyear{FWR1}; Wagenmakers, Farrell, \& Ratcliff, \citeyear{WFR1,WFR2}; Wagenmakers, van der Maas, \& Farrell \citeyear{WvF}). In an alternative nomothetic perspective (Dixon, Holden, Mirman, \& Stephen, \citeyear{DHMS}; Dixon, Stephen, Boncoddo, \& Anastas, \citeyear{DSBA}; Gilden, \citeyear{Gil01}; Gilden et al., \citeyear{GiTM}; Holden, \citeyear{Hol13}; Ihlen \& Vereijken, \citeyear{IV1}; Riley \& Holden, \citeyear{RiHo}; Stephen, Boncoddo, Magnuson, \& Dixon, \citeyear{SBMD}; Stephen, Dixon, \& Isenhower, \citeyear{SDI}; Van Orden et al., \citeyear{VOHT1}), this stochastic behavior may be due not to specific cognitive systems, nor to the mere sum of their noises, but to a more fundamental mechanism such that cognition would happen as the emergence of patterns in a self-organizing complex dynamical system. In particular, the $1/f$ scaling might be the collective expression of the metastable coordination of different cognitive and motor systems in the performance of a task \citep{KBHVO}.\footnote{Criticism to the nomothetic view can be found in \citet{FWR1} and Wagenmakers et al. \citeyearpar{WFR1,WvF}; early replies are by Thornton and Gilden \citeyearpar{TG} and \citet{VOHT2}, while a more recent defense is by Ihlen and Vereijken \citeyearpar{IV2}. A somewhat intermediate view between the idiosyncratic and the nomothetic was proposed by Likens, Fine, Amazeen, and Amazeen \citeyearpar{LFAA}. Other thoughts about the intrinsic uncertainty of the $1/f$-scaling phenomenon and the role of measurement in psychological experiments, related by analogy with quantum physics, can be found in Holden, Choi, Amazeen, and Van Orden \citeyearpar{HCAV} and Van Orden, Kello, and Holden \citeyearpar{VOKH}.}

Typically, the variation of the response as a function of the trial is, when plotted over thousands of trials, a highly rugged (more precisely, nowhere differentiable) curve. This curve, or graph, has the same mathematical properties of certain stochastic processes found in statistical mechanics and anomalous transport theory \citep{MeK04,Sok12}. These processes are self-similar in a probabilistic sense and are naturally associated with random fractals. The frequency distribution $P(f)$ is the generalization of the number of contraction maps defining deterministic fractals. In this case, one has a random fractal and $P$ has a spectrum of values not distributed in the integer field. The ``fractal dimension'' is an ambiguous concept here, that depends on whether one is considering the walk of the stochastic process (vertices and edges may repeat), the trail (vertices may repeat, edges do not), the path or graph (vertices and edges do not repeat), the set of zeros of the path, and so on. In cognitive psychology, one usually refers to the trial series. To be precise, there are two stochastic processes of interest that can closely describe the typical trial series. One is fractional Brownian motion (FBM) \citep{BaA,MaV} and the ``fractal dimension'' is the Hausdorff dimension of its graph $\cG$. Fractional Brownian motions are characterized by a parameter $0\leq H< 1$ called Hurst exponent, and they produce a spectrum with $\de=2H+1$. The Hausdorff dimension of $\cG$ is equal to the box-counting dimension for these stochastic processes and reads $\dh=d_\textsc{b}=2-H$ almost surely, i.e., with probability 1 \citep{Fal03}. Different values of $\dh$ are associated with various ``noises'' and frequency distributions $P(f)\sim f^{2\dh-5}$, ranging from $\dh=2$ [$H=0$, $P(f)\sim f^{-1}$, ideal pink or flicker noise] to $3/2<\dh<2$ [$0<H<1/2$, $P(f)\sim f^{-1}-f^{-2}$, antipersistent FBM], the special case $\dh=3/2$ [$H=1/2$, $P(f)\sim f^{-2}$, Wiener process (aka standard Brownian motion), no correlation of increments], and $1<\dh<3/2$ [$1/2<H<1$, $P(f)\sim f^{-2}-f^{-3}$, persistent FBM].\footnote{See \citet{EHKK} and \citet{Hol05} for this classification of $P(f)$ in psychology and \citet{Fal03} for the proof that $\dh=2-H$ almost surely for standard and fractional Brownian motion. Still in \citet{Fal03}, also the Hausdorff dimension of the trail of $D$-dimensional Brownian motion is calculated and is $\dh=2$ almost surely for $D\geq 2$.} The other process of relevance triggers for frequency distributions with $-1<\de=2H_\textsc{fgn}-1<1$, which are described by a fractional Gaussian noise with Hurst exponent $0<H_\textsc{fgn}<1$ \citep{EHKK}. By analytic continuation of the expression of $\de$, one can identify $H=H_\textsc{fgn}-1$ and define the fractal dimension (with no further specification) as $d=3-H_\textsc{fgn}$ so that $P(f)\sim f^{2d-5}$. Apart from the $d=2$ case giving ideal pink noise ($H_\textsc{fgn}=1$), the three main regimes are $2<d<5/2$ [$1/2<H_\textsc{fgn}<1$, $P(f)\sim f^0-f^{-1}$, nonideal pink noise, characterized by small fluctuations at short time scales and larger fluctuations modulated on longer time spans], $d=5/2$ [$H_\textsc{fgn}=1/2$, $P(f)\sim f^0$, white noise, equally sized fluctuations with no time correlation], and $5/2<d<3$ [$0<H_\textsc{fgn}<1/2$, $P(f)\sim f^0-f^1$, nonideal blue noise]. The value of $\dh$ or $d$ can change according to the experimental conditions and participants but, as said above, it reproduces nonideal or almost ideal pink noise. Thus, human variability in the performance of a task can be described by a random fractal of dimension $2\lesssim d<2.5$.

The initial idea when $1/f$ cognitive noise was discovered was that the internal biological clock (in memory tasks) and other cognitive systems involved in reaction tasks generate a $1/f$ (``ideal pink'') noise, while the motor system and the experimental design produce a white-noise interference [a horizontal line in the $(\log f,\log N)$ plane] \citep{GiTM}. Subtle changes in task demands introduce an exogenous variation in the performance and modify the spectral distribution $P(f)$ [the line with slope $-\de$ in the $(\log f,\log N)$ plane] representing what is interpreted as the endogenous (or fundamental, intrinsic to mind and body) variation. To reproduce the observed deviation from a straight line, it was proposed to look for white noise in data, which would flatten the $1/f$ noise line at high frequencies \citep{GiTM}. However, experiments carried with humans failed to confirm this ``layering hypothesis'' \citep{HCAV,IV1}, pointing instead towards a \emph{multifractal} noise with different exponents $\de$ at different scales (Dixon et al., \citeyear{DHMS,DSBA}; Holden et al., \citeyear{HCAV}; Ihlen \& Vereijken, \citeyear{IV1,IV2}; Stephen, Boncoddo, et al., \citeyear{SBMD}; Stephen \& Dixon, \citeyear{StDi}; Stephen, Dixon, et al., \citeyear{SDI}).\footnote{A psychologist-oriented review on the concepts of multifractals and multiplicative cascades, which are special multifractal distributions of points, is by Kelty-Stephen, Palatinus, Saltzman, and Dixon \citeyearpar{KSPSD}. See also Nonaka and Bril \citeyearpar{NB} for an example of ``multifractal'' performance.}

Let us now go back to associative models and see whether there is some relation between the fractal structure we found and that of $1/f$-scaling cognitive scenarios. The most conservative view is that there is no connection at all, for several reasons. First, associative models describe behavior in Pavlovian conditioning, while the $1/f$-noise effect is found in very different cognitive tasks. Second, although the paradigm of behaviorism states that the rules of human behavior can be inferred from those of animal behavior, strictly speaking the established range of applicability of associative models does not overlap with the experiments in human response. The third reason, encompassing the other two, is that trying to associate similar mathematical structures arising in different contexts may be dangerous if there is no principle guiding us, apart that of cursory resemblance. 
 In general, one can regard associative models of learning as useful tools without making any claim on their validity as bits of a more fundamental theory of the human mind.

\begin{figure}
\centering
\includegraphics[width=7.8cm]{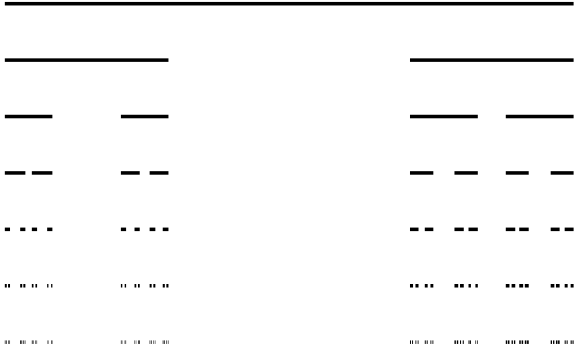}
\caption{\label{fig6} A random Cantor set $C_\la$ corresponding to a randomized Hull model of the second type for $n=6$ iterations, with $\la=1$ and $\a\b$ picking values in the interval $(2/3-0.1,2/3+0.1)$ with uniform probability distribution. In this example, the CS and US saliences vary at each trial.}
\end{figure}

Having said that, under a more optimistic perspective, cognitive and behavioral psychology must agree to some extent, as they both approach the same topic (animal and human conduct) albeit from different directions (internalist versus environmentalist). In a preliminary attempt to make this link, and without the pretension of being rigorous, we notice that the random fractal structures found in cognitive experiments can only be compared with the randomized version of our proposal, not with the deterministic one of the previous sections. Consider a random Cantor set obtained by varying any of the parameters $\la$, $\a$, and $\b$ randomly in the interval $(0,1)$. To replicate as much as we can the typical situation of the cognitive experiments, we cannot change the intensity of the US, which must remain the same at all trials. What can change is the combined CS--US salience $\a\b$. We will call such random Cantor set $C_\la$. We can imagine that the random variation of $\a\b$ at each trial is due to the same internal mechanisms of the $1/f$ noise, let them be the superposition of various cognitive systems or the emergence of a complex pattern. In Fig.\ \ref{fig6}, we plot a random Cantor set with just the desired features: $\la=1$ is fixed while $\a\b$ can take a random value (with uniform distribution) in the interval $(2/3-0.1,2/3+0.1)$. The central value $\a\b=2/3$ corresponds to a deterministic ternary Cantor set and the maximal fluctuation $\pm 0.1$ (unrealistically large in order to show the effect pictorially) represents (in a cognitive interpretation) the putative internal cognitive noise. It is useful to make a comparison with Mackintosh model, where $\a$ changes with each trial but deterministically. Both in that case and in our random model, the cognitive process affecting conditioning is attention (to the CS). However, while in Mackintosh model $\a$ has a deterministic gradient $|\a_{\rm max}-\a_{\rm min}|$ throughout the duration of the experiment (deterministic because determined completely by the way attention to the CS increases or decreases as the level of new information it carries changes), here it suffers small random variations due to the internal flickering of the attentional system (in the idiosyncratic view) or the global internal flickering from the interaction of the attentional and other cognitive systems (in the nomothetic view).

Comparing the geometric properties of the fractal $C_\la$ with those of the stochastic graph $\cG$ discussed above may be a tricky issue, since they represent different things in different types of experiments. In the first case, the fractal $C_\la$ is a set of points corresponding to all possible values of what can be measured for all possible initial conditions $0\leq V_1\leq \la$, i.e., a suitably operationalized internal variable representing the CS--US association strength of a subject in a Pavlovian conditioning experiment. In the second case, the fractal $\cG$ is the actually measured series of the responses of a subject in a cognitive task. However, what \emph{could} be interesting to check would be, on one hand, whether the random model $C_\la$ fits data better than the deterministic one $\cC$ and, on the other hand, in an experiment uniting the features of both Pavlovian conditioning and $1/f$ cognitive tasks, whether its application to human subjects could capture a modulation in their response that could be statistically related to the modulation found in cognitive psychology. We leave the verification of this possibility to future studies. For the time being, we cannot help but notice an intriguing parallelism: the Hausdorff dimension of the fractals associated with conditioning processes decreases (from 1 to 0) with the efficiency of conditioning, while the Hausdorff or ``fractal'' dimension of the stochastic response pattern in cognitive tasks decreases (to values $\gtrsim 2$) with the improvement of the performance (Castillo, Kloos, Holden, \& Richardson, \citeyear{CKHR}). In the first case, a better performance means a faster conditioning and a more dust-like set, while in the second case a better performance literally means a smoother performance. The quantitative theoretical description proposed here yields cautious support to what found experimentally in cognitive science and to the notion that measurable behavior can be characterized, in a precise sense, by irregular geometry.



\section{Implications of the theory}\label{conseq}

Describing the geometric properties of traditional conditioning models in terms of fractals may be a worthwhile mathematical exercise, but its real value to the discipline should be measured in terms of its practical applications. We mention two, one theoretical and one experimental.

The theoretical application pertains to the deterministic and random version of the theory alike, and is a new, or different, understanding of the psychological process underlying conditioning. Consider the Hull model with one CS. The learning rate is determined by the product $\a\b$ of the CS and US saliences in an intuitive way: when the salience of the stimuli is low, the subject takes longer (i.e., more trials, more sessions) to acquire maximal association between the stimuli. We have seen that the same statement can be recast in terms of the dimension of the fractal associated with the model: low saliences correspond to a set ``spilling over the line'' with Hausdorff dimension $\dh>1$, while high saliences correspond to an ordinary dust-like, totally disconnected Cantor set with $\dh<1$. From this reformulation, we can gain a deeper insight into the nature of the process (learning) this model attempts to describe. Since faster learning is associated with a sparsely populated, totally disconnected set of points with $\dh<1$, it can be regarded as a process making large ``hops'' between points in the support of the fractal. This happens because the interval $I=[0,\la]$ at the zeroth iteration is depleted of points faster at each iteration. Another consequence of this low dimensionality is that the excitatory and inhibitory branches have no mutual intersection and each point on the line is uniquely associated with only one similarity branch. In a sense, excitatory conditioning is uncontaminated by the inhibitory one. On the other hand, when $\a\b$ is small and $\dh>1$, the inhibitory and excitatory branches share points and this superposition of otherwise separated processes gives rise, through a sort of contamination by the inhibitory branch, to slow learning. Of course, the acquisition rate is not an intrinsic quality of a process but a subject-dependent feature. At this point, the theory might naturally open up the possibility of the existence of cognitive interference in the internal workings of a slow-learning subject, but we prefer to leave further speculations to the curious reader.

\begin{figure}
\centering
\includegraphics[width=7.8cm,height=7.5cm]{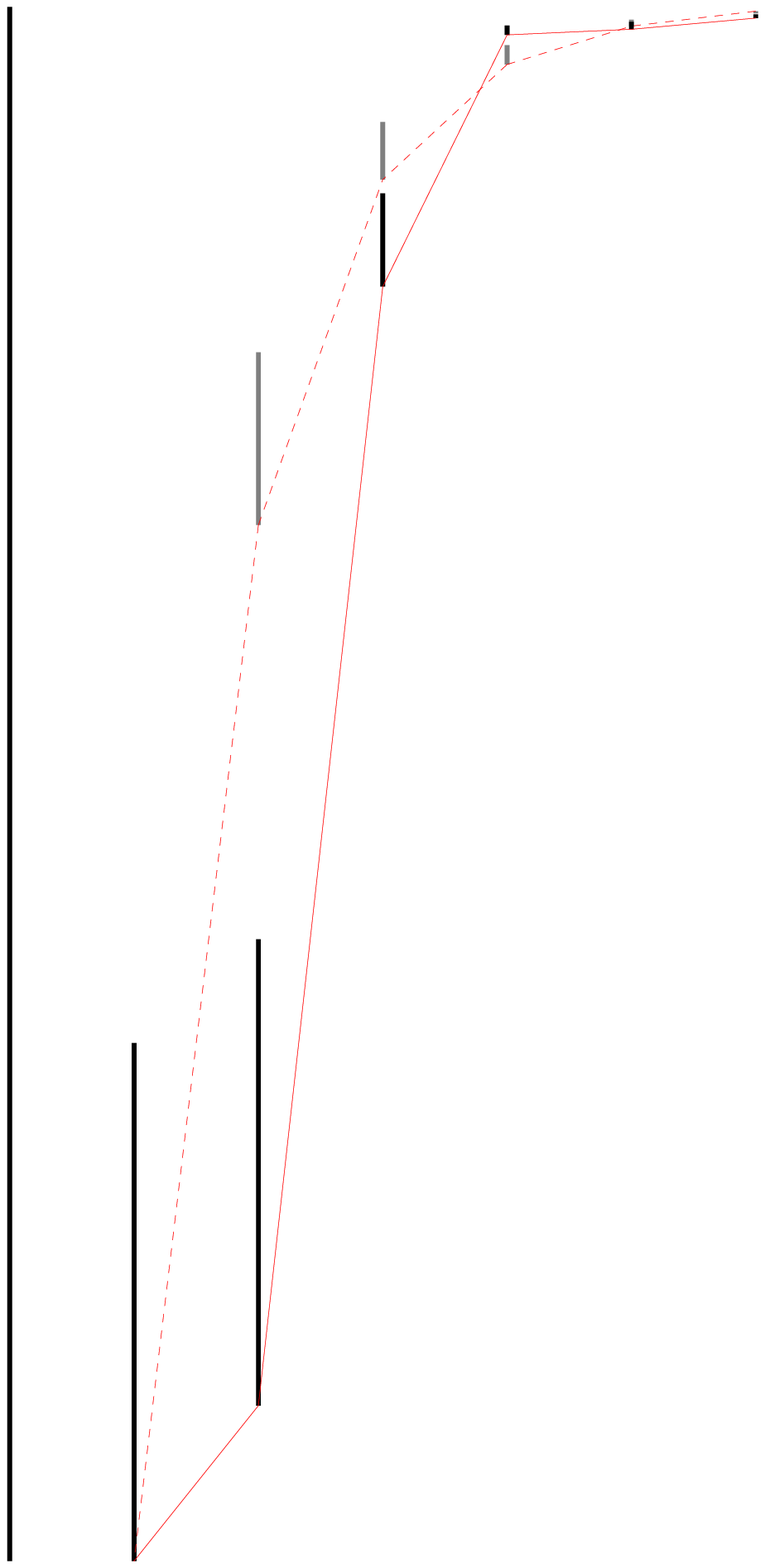}\\
\includegraphics[width=7.8cm,height=7.5cm]{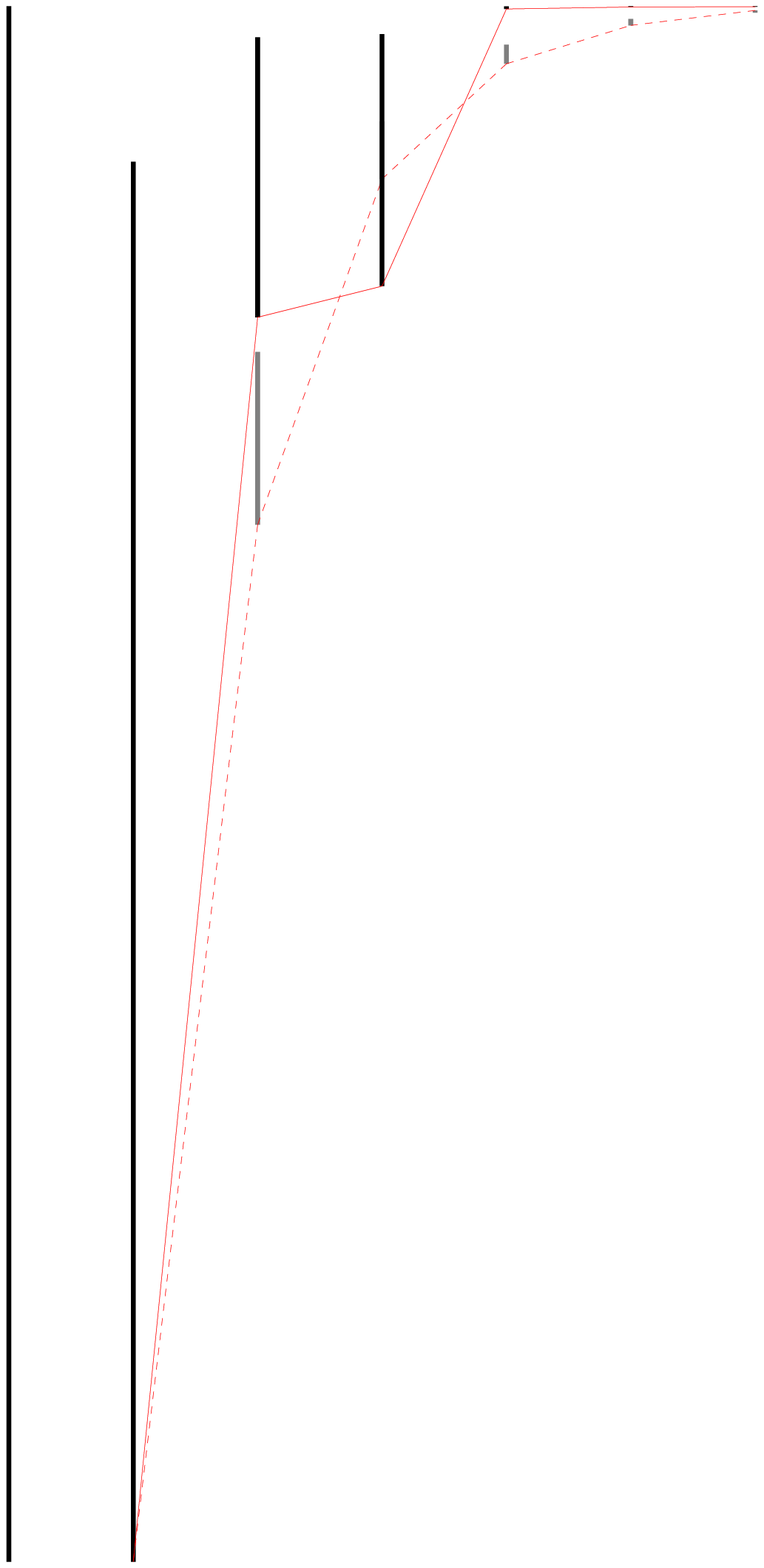}
\caption{\label{fig7} Examples of excitation learning curves (solid red) for random Cantor sets corresponding to a randomized Hull model for $n=6$ iterations (from left to right), with $\la=1=\b$ and $\a$ picking values in the interval $(0,1)$. Top panel: $\a_1=1/3$, $\a_2=0.1$, $\a_3=0.8$, $\a_4=0.9$, $\a_5=0.2$, $\a_6=0.5$. Bottom panel: $\a_1=0.1$, $\a_2=0.8$, $\a_3=0.1$, $\a_4=0.99$, $\a_5=0.7$, $\a_6=0.4$. Black segments are the portions of the Cantor set where the learning curve touches upon. The dashed red curve corresponds to the deterministic curve with $\a=1/3$ (ternary Cantor set, gray segments).}
\end{figure}

The experimental application belongs only to the random version of the theory. The deterministic version is mathematically equivalent to traditional associative models and their predictions are the deterministic-fractal picture's predictions. However, the random version of the theory discussed in Section \ref{relp} does make a forecast, unreachable without interpreting associative models as fractals, about the pattern of behavioral variations in the case of a randomly varying stimulus salience or magnitude. Consider the case where the saliences $\a\b$ and the magnitude $\la$ of the US take a random value in a given distribution with support between 0 and 1. As we have seen, this can happen in different situations, from a controlled experimental design of partial reinforcement with randomized schedule (the CS or the US can be either present or absent at any given trial; then, $\la$ takes the values 0 or 1) to the natural environment of the subject with everchanging stimuli (where $\la$ can take different values at each trial). Or, according to the hypotheses put forward in the $1/f$ cognitive literature, random variations of $\a\b$ may happen due to the internal flickering of the subject's cognitive modules. We can distinguish between two general effects of the variation of these parameters: monotonic effects and fluctuations in the subject response.
\begin{itemize}
\item When $\la$ takes only two values (0 or 1, absence or presence of the US) and $\a$ is fixed, the actual learning curve is monotonic and systematically \emph{below} the absolute asymptote of learning $\la$ of the deterministic theoretical model (Fig.\ \ref{fig4}, first type of randomized Cantor set, where $\b$ is also fixed). This is a simple consequence of the fact that, if $\a\b>0$ (positive US and CS saliences), then all subsequent copies of $I$ are equal to or smaller than the one in the previous iteration. If $\b$ is smaller at trials where $\la=0$, then $\a\b$ is smaller, the scaling ratio $1-\a\b$ becomes larger, the shift $(1-\a\b)\la$ in the mapping $\cS_1^n$ becomes larger, the image $\cS_1^n\circ\cS_2(I)$ is shifted more to the right, and the effect is smaller. Probably, this effect can be checked only with averaged data, since it could be masked by fluctuations caused by individual differences. A carefully controlled partial reinforcement program can achieve this if the schedule of presentation of the US is the same for all experimental subjects.
\item When $\la$ and $\b$ are fixed (continuous reinforcement) and the CS salience varies randomly in the interval $0<\a<1$, then the association strength can be either \emph{above or below} the theoretical point of the deterministic model (with parameters $\la$, $\b$ and $\a_{\rm average}$) at a given iteration, depending on how $\a$ varies. This creates a pattern of fluctuations around the deterministic theoretical curve (Fig.\ \ref{fig7}). Of course, the experimenter cannot control the random variation of the CS salience if this depends on some variable internal to the subject. However, one can formulate an expectation of the general trend of response variability. Since the support of the random fractal is a subset of the interval $I=[0,\la]$ and all shifts are smaller than $\la$, points lie within this range while points $V>\la$ do not belong to the geometric construction of the process. Therefore, fluctuations in the subject response is predicted to be either as large as $O(\la)$ but asymmetric (mostly \emph{below} the theoretical deterministic learning curve) or symmetric around the curve but relatively \emph{small}. In both cases, fluctuations are progressively and quickly \emph{damped} as the curve approaches the asymptote. Contrary to the previous case, these features should be looked for in individual data, since damped small response fluctuations would be easily flattened in averaged data.
\item When also $\la$ varies randomly, as in a natural setting or in the laboratory when the US magnitude is changed at every trial, also in this case the learning curve of individuals is characterized by a certain variability. Fluctuations above or below the deterministic theoretical curve follow about the same pattern just described and they are tuned by multiple products of the scaling ratios $1-\a_n\b_n$ and the US magnitude $\la_n$.
\end{itemize}


\section{Conclusions}\label{concl}

Eventually, single-variable linear or nonlinear equations cannot account for the variety of conditioning processes we are aware of. Systems of coupled iterative laws with multiple entangled variables are typical in modern approaches such as the SOP model of memory processing (Brandon et al., \citeyear{BVW}; Wagner, \citeyear{Wag81}), where elements of different nodes in a neural graph interact nontrivially. Sensitivity to context further complicates the way different stimuli interact, as reflected by later elemental theories \citep{BW,Wag03,WB} [see \citet{Wag03} and the crystal-clear reviews by \citet{Wag08} and Wagner and Vogel \citeyearpar{WV}, also for an account on configural theories]. Nevertheless, simple models such as Rescorla--Wagner and Pearce--Hall have not exhausted their usefulness. For instance, they are still topical in as hot a field of research as neuroscience and they may actually coexist in models of error signal processing in the brain \citep{RELDS}. Furthermore, our fractal approach resonates in some yet unfathomable but intriguing way with the findings on task performance in cognitive psychology, as discussed in Section \ref{relp}. The extension to multifractals will be a most natural direction where to look into, since it could connect with the analogous multiscale phenomena met in cognitive experiments.

All this leads us to believe that the examples we presented here are not just foundational prototypes of a more involved paradigm. The alternative toolbox of fractal geometry, of which we saw examples in the Hausdorff dimension as a means to rank the efficiency of conditioning from the subject-environment interaction, and in random fractals as descriptions of a variety of programs (including of partial conditioning), or even of conditioning with variable performance due to internal biological fluctuations, may already lend itself to promising applications.

\paragraph*{Acknowledgments} The author is under a Ram\'on y Cajal contract and thanks Ricardo Pellón for guidance and useful discussions, and John G.\ Holden for very helpful comments on the manuscript.

\bibliographystyle{apacite}

\end{document}